\documentclass[11pt,a4paper]{article}

\usepackage{jheppub}
\usepackage[svgnames]{xcolor}
\usepackage{tikz}
\tikzset{>=latex}
\tikzset{baseline=(A.base)}
\tikzset{arbitrary/.style={}}
\tikzset{1gon/.style={draw}}
\usetikzlibrary{shapes.geometric}
\tikzset{SU/.style={draw,circle}}
\tikzset{Sp/.style={draw,semicircle,inner sep=2pt,shape border rotate=180}}
\tikzset{SO/.style={draw,semicircle,inner sep=2pt}}

\def\SO{\mathrm{SO}}
\def\SU{\mathrm{SU}}
\def\USp{\mathrm{USp}}
\def\U{\mathrm{U}}

\def\cN{\mathcal{N}}

\def\fund{\mathbf{fund}}
\def\vect{\mathbf{vect}}
\def\adj{\mathbf{adj}}
\def\asym{\mathbf{asym}}
\def\asymT{\mathbf{asym3}}
\def\sym{\mathbf{sym}}
\def\spinor{\mathbf{S}}
\def\conj{\mathbf{C}}
\def\half{\mathbf{\tfrac12}}

\title{Classification of 4d $\cN{=}2$ gauge theories}

\preprint{IPMU-13-0179, UT-13-33}

\author[1]{Lakshya Bhardwaj}
\affiliation[1]{Department of Physics,  Chennai Mathematical Institute,\\
Plot H1, SIPCOT IT Park  Siruseri, Kelambakkam 603103, India}

\author[2]{and Yuji Tachikawa}
\affiliation[2]{Department of Physics, Faculty of Science, \\
 University of Tokyo,  Bunkyo-ku, Tokyo 133-0022, Japan and \\
Institute for the Physics and Mathematics of the Universe, \\
 University of Tokyo,  Kashiwa, Chiba 277-8583, Japan}

\abstract{
We classify all possible four-dimensional $\cN{=}2$ supersymmetric UV-complete gauge theories composed of semi-simple gauge groups and hypermultiplets. 
We also give appropriate references for all theories with known Seiberg-Witten solutions.
}

\keywords{}

\begin{document}
\setcounter{tocdepth}{2}
\maketitle
\section{Introduction}
\begin{figure}
\centering
\begin{tikzpicture}[anchor=mid]
\filldraw[draw=PaleVioletRed!75!black,fill=PaleVioletRed,fill opacity=0.5] (.8,0) circle (1);
\filldraw[draw=Gold!75!black,fill=Gold,fill opacity=0.5] (2.8,0) circle (2.3);
\draw (-1.4,0) node (A) {\parbox{3.5em}{Lagrangian \\ gauge  theories}};
\draw[->,>=latex] (A)--(.25,0);
\draw (3.3,0) node {Class S theories};
\draw[black,rounded corners=1cm] (-2.5,3.5) rectangle (5.5,-2.5);
\draw (1.5,2.8) node {$\cN{=}2$ supersymmetric theories};
\end{tikzpicture}
\caption{A subjective view of $\cN{=}2$ supersymmetric theories\label{subjective}}
\end{figure}
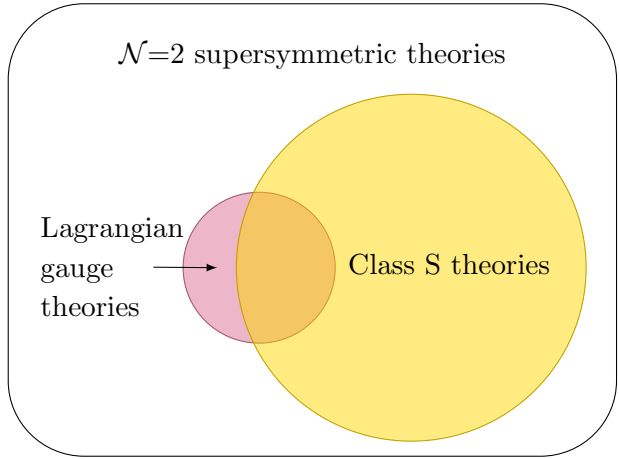

Four-dimensional $\cN{=}2$ supersymmetric gauge theories, first introduced in the 1970s \cite{Fayet:1975yi,Brink:1976bc,Grimm:1977xp,Fayet:1978ig}, have been a fertile playground for mathematical physicists for the last 20 years. 
First came the seminal works of Seiberg and Witten \cite{Seiberg:1994rs,Seiberg:1994aj}, where the infrared dynamics of $\SU(2)$ gauge theory was exactly determined up to two derivatives. 
After ten years of activity extending their analysis to various other groups and matter contents, Nekrasov showed how to derive their solutions from an honest path integral computation \cite{Nekrasov:2002qd}. 
Again, this method was slowly generalized to other gauge groups and hypermultiplet representations. 
A new impetus to the study of $\cN{=}2$ dynamics was given by Argyres and Seiberg \cite{Argyres:2007cn} and Gaiotto \cite{Gaiotto:2009we}, where it was shown that in order to fully understand the S-duality properties of $\cN{=}2$ supersymmetric theories it is essential to consider as first-class citizens  the theories which consist of strongly-coupled superconformal field theories further coupled to gauge fields. 
Many of these theories can be understood as a compactification on a Riemann surface of the six-dimensional $\cN{=}(2,0)$ theories, and they are recently called theories of class S. See Fig.~\ref{subjective} for a subjective view of the $\cN{=}2$ theories by the authors. 

Still,  $\cN{=}2$ supersymmetric UV-complete gauge theories composed of gauge groups and hypermultiplets form a nice traditional subclass of all possible four-dimensional $\cN{=}2$ supersymmetric systems. The main aim of this paper is to classify all such traditional theories. There are a few rationales behind such a classification.
\begin{itemize}
\item The classification of $\cN{=}4$ supersymmetric gauge theories is equal to the classification of semi-simple groups, and therefore is trivial in a sense.\footnote{Here we do not consider possible mysterious non-Lagrangian $\cN{=}4$ theories. For these, see e.g.~\cite{Beem:2013qxa}.} The classification of $\cN{=}1$ supersymmetric or non-supersymmetric gauge theories is surely not possible in any explicit manner, because the classification of superpotentials or potentials are involved. 
The classification of $\cN{=}2$ gauge theories lies in the middle of these two extremes, and is actually possible. 
\item The classification of theories consisting of $\SU$ gauge groups and bifundamentals, is known to be equivalent to the classification of affine and non-affine Dynkin diagrams \cite{Katz:1997eq}.   One might expect that the full classification would also have comparably nice structures.
\item In the last 20 years, people have developed many mutually complementary methods, to obtain the Seiberg-Witten solutions of $\cN{=}2$ gauge theories. However, none of them applies to arbitrary $\cN{=}2$ gauge theories, and each of the methods have strengths and weaknesses depending on the type of gauge theories. It would be interesting to know if the methods obtained so far are powerful enough to solve all $\cN{=}2$ UV-complete gauge theories, and if not, to develop new methods to solve them. 
\end{itemize}

The result of the classification we will carry out can be stated as follows. 
First, in this paper we do not distinguish gauge groups with the same Lie algebra.
Given two $\cN{=}2$ UV complete gauge theories, we can consider a new gauge theory which is just a direct sum of the two decoupled systems. We call such a trivially combined  theory reducible. Below we only discuss irreducible theories, which we find to  fall into one of the three possibilities: \begin{enumerate}
\item The gauge group is simple. This was classified in \cite{Koh:1983ir,Dong:1984vt,Derendinger:1984bu}.
\item The gauge group is $\SU(2)^n$ and the hypermultiplets are given by trifundamentals, first introduced by Gaiotto \cite{Gaiotto:2009we}.
\item The gauge group is of the form $G=\prod_i G_i$ with $G_i$ simple or $\SO(4)$, and each irreducible component of the hypermultiplets is charged under at most two $G_i$. We can then associate a graph where each $G_i$ corresponds to a node and where each irreducible hypermultiplet charged under $G_i$ and $G_j$ corresponds to an edge connecting two nodes.  By a simple argument we will find that the associated graph is just a single loop or a tree. In the case of a tree, it can always be built by attaching a number of \emph{branches} to a single \emph{trunk}.  The possible types of branches and trunks are classified, and the allowed ways of attaching branches to a trunk are enumerated. 
In the end we will find that there are only a finite number of gauge theories whose associated graph is neither a finite nor an affine Dynkin diagram. 
\end{enumerate}

We give an explicit and straightforward algorithm to enumerate all gauge theories, and a Mathematica file implementing that algorithm accompanies this paper as an ancillary file. 
We also give a more detailed summary of the classification in Sec.~\ref{summary}.

For each theory, we will also note whether Seiberg-Witten solution is already known or not in July 2017,
and a reference to the literature if a solution is publicly available.\footnote{The authors apologize in advance that they only tried to list at least one paper for each solved theory, and did not try to be exhaustive in listing the original paper. They will appreciate inputs from the readers, and will be happy to cite more papers. } There are four major complementary methods to obtain the Seiberg-Witten solutions: \begin{itemize}
\item The first is to come up with a good integrable system and then to show that it works \cite{Gorsky:1995zq,Donagi:1995cf,Martinec:1995by,Itoyama:1995nv,D'Hoker:1998yi}.
\item The second is to compactify  6d $\cN{=}(2,0)$ theory on a Riemann surface, possibly with punctures \cite{Klemm:1996bj,Witten:1997sc,Evans:1997hk,Landsteiner:1997vd,Brandhuber:1997cc,Landsteiner:1997ei,Kapustin:1998xn,Landsteiner:1998pb,Argyres:2002xc,Gaiotto:2009we,Nanopoulos:2009xe,Chacaltana:2010ks,Tachikawa:2010vg,Chacaltana:2011ze,Chacaltana:2012ch,Chacaltana:2013oka}.
\item The third is to find a local Calabi-Yau on which we compactify the type II string theory \cite{Kachru:1995wm,Katz:1996fh,Katz:1997eq,Brodie:1997qg,Aganagic:1997wp,Terashima:1998iz,Terashima:1998fx,Hashiba:1999ss,Tachikawa:2011yr}.
\item The fourth is to perform the instanton counting \cite{Nekrasov:2002qd,Nekrasov:2003rj,Nekrasov:2004vw,Marino:2004cn,Shadchin:2005cc,Hollands:2010xa,Hollands:2011zc,Nekrasov:2012xe}.
\end{itemize}
It is sometimes the case that some of the theories we find during the classification are clearly solvable with one of the methods list above but that the solution is not yet available in the literature. We also comment on them briefly. 

The rest of the paper is organized as follows.
In Sec.~\ref{definingdata}, we summarize the defining data of $\cN{=}2$ supersymmetric gauge theories.
Then in Sec.~\ref{easypart}, we first outline our method of classification. We then show that any UV-complete theory falls within the three-way distinctions given above, and carry out the classification of the first two classes. 
In Sec.~\ref{hardpart}, we perform the classification of the third class of theories. We first perform a local analysis at each node, and show that the associated graph is either a single loop or a tree. We then classify all the possible trunks and the branches. 
In Sec.~\ref{summary} we conclude our paper by summarizing the result of the classification.

\section{Defining data of $\cN{=}2$ gauge theories}\label{definingdata}
Let us first recall the structure of possibly non-UV-complete $\cN{=}2$ supersymmetric gauge theories.
An $\cN{=}2$ supersymmetric gauge theory is specified by its gauge group and its hypermultiplets.
The gauge group is given by  \begin{equation}
G=G_1\times G_2\times \cdots G_s,
\end{equation} where $G_i$ is either simple or $\U(1)$.  In this paper we only distinguish the gauge sector up to its Lie algebra, and do not discuss its global structure and/or subtle theta angles associated to the choice of the line operators \cite{Gaiotto:2010be,Aharony:2013hda}.

As for hypermultiplets, the basic ingredient is a half-hypermultiplet in a pseudo-real representation $R$. This consists of an $\cN{=}1$ chiral multiplet in the representation $R$. For a complex representation $R$, there is a standard pseudo-real structure on $R\oplus \bar R$. Then a half-hypermultiplet in $R\oplus \bar R$ is called a (full) hypermultiplet in the representation $R$. This consists of a pair of $\cN{=}1$ chiral multiplets, one in the representation $R$ and another in the complex conjugate representation $\bar R$. Half-hypermultiplets were first introduced in \cite{Ferrara:1981ep,deWit:1983rz}.

Let us say that we have half hypermultiplets in the representation \begin{equation}
\bigoplus_a R_a 
\end{equation} where 
\begin{equation}
R_a=R_{1,a}\otimes R_{2,a} \otimes \cdots \otimes R_{s,a}
\end{equation}
with $R_{i,a}$ being an irreducible complex representation of $G_i$. $
R_{i,a}$ can be a trivial one-dimensional representation. 
When $R_a$  is pseudo-real, it can appear alone.
This is equivalent to the condition that an odd number of $R_{i,a}$ are pseudo-real and other $R_{i,a}$ are strictly real.
Otherwise, there should be another index $b\neq a$ such that $R_{a}=\bar R_{b}$ to form a full hypermultiplet.  
Thus we can organize the hypermultiplet representations as follows: \begin{equation}
\left[\bigoplus_x R_{x}\right] \oplus \left[\bigoplus_y  R_y \oplus \bar R_y \right]\label{matter}
\end{equation} where $x$ runs over proper half-hypermultiplets
and $y$ runs over full hypermultiplets. 
For each $y$, we can turn on $\cN{=}2$ supersymmetric mass terms. 
As it does not affect the UV behavior of the theories, we do not discuss them further, unless absolutely necessary. 

The coefficient of the one-loop beta function of $G_i$ is given by \begin{equation}
b_i := 2h^\vee(G_i) - \sum_a c_2(R_{i,a}) \prod_{j\neq i} \dim R_{j,a} 
\label{beta0}
\end{equation} where $h^\vee(G_i)$ is the dual Coxeter number of $G_i$ (which we take to be zero for $G_i=\U(1)$), and $c_2(R)$ is often called the index of the irreducible representation $R$.
The UV-completeness of the theory requires that $b_i$ for all $i$ is non-negative. 
For this gauge theory to be UV-complete, $G_i$ cannot be $\U(1)$, and therefore $G$ is semisimple.

When $G_i$ is $\USp(2n)$, we need to make sure that there is no global anomaly \cite{Witten:1982fp}, independent of whether the theory is UV-complete or not.  This is equivalent to the condition that $b_i$ is integer. 

For our purposes, it is more convenient to re-organize \eqref{beta0} according to the contribution from half-hypermultiplets
and full hypermultiplets in \eqref{matter}: \begin{equation}
b_i=2h^\vee(G_i)-\sum_x b_i(R_x) - \sum_y b_i(R_y)\label{beta}
\end{equation} where \begin{equation}
b_i(R_x)=c_2(R_{i,x}) \prod_{j\neq i} \dim R_{j,x}
\end{equation} for a half-hypermultiplet and \begin{equation}
b_i(R_y)=2c_2(R_{i,y}) \prod_{j\neq i} \dim R_{j,y}
\end{equation} for a full hypermultiplet. 
We call $b_i(R)$ as the contribution to the one-loop beta function of $G_i$ from the (half or full) hypermultiplet in $R$.
We also sometimes use the notation $b_{G_i}(R)$ to denote the same quantity.

We can summarize the data defining an $\cN{=}2$ gauge theory in a following combinatorial object: 
\begin{itemize}
\item For each $G_i$, associate a node labeled by $G_i$.
\item For each half-hypermultiplet $R_x$ in \eqref{matter},
let  $n$ be the number of non-trivial representations $R_{i,x}$.
We then draw an $n$-gon labeled by $R_x$ 
with its $n$-vertices given by the nodes $G_i$
such that $R_{i,x}$ is non-trivial. 
\item For each full hypermultiplet $R_y\oplus \bar R_y$ in \eqref{matter},
we draw an $n$-gon as above, based on $R_y$, and label it with $R_y \oplus \bar R_y$. 
\item We of course need to distinguish a node $G_i$ and a 1-gon at a node $G_i$ labeled with a representation $R_a=R_{i,a}$. In practice we draw a 2-gon with only one end attached to a node instead. 
A 2-gon is regarded as an edge. 
\end{itemize}
For a UV-complete theory, we require that at each node, the quantity \eqref{beta} is non-negative and integral.

We call this combinatorial object the associated graph of the $\cN{=}2$ theory. 
The $\cN{=}2$ theory, up to the choice of the complexified coupling constants and the mass terms, can be reconstructed from its associated graph. Therefore, for the purpose of this paper we identify a theory and its associated graph.

\section{The classification, part I}\label{easypart}
The classification consists of  the following steps, which we go through in turn: \begin{itemize}
\item[Sec.~\ref{nodes}:]
 We list all possible nodes of the associated graph, or equivalently, all possible simple factors of the gauge group.
\item[Sec.~\ref{ngons}:] 
We list all possible $n$-gons of the associated graph, or equivalently, all possible irreducible half or full hypermultiplets. We will find that there are no $4$-gons or higher, and that the available 3-gons are the trifundamental of either $\SU(2)^3$ or $\SU(2)^2\times \USp(4)$.  We will allow a node of the associated graph to represent $\SO(4)$, and regard the latter as a $2$-gon connecting an $\SO(4)$ node and a $\USp(4)$ node.
\item[Sec.~\ref{reduction}:] 
We show that the classification of asymptotically-free gauge theories can be reduced to the classification of superconformal gauge theories.
\item[Sec.~\ref{gaiotto}:] 
We show that any theory containing at least one $\SU(2)^3$ trifundamental is an $\cN{=}2$ gauge  theory of the type introduced by Gaiotto in \cite{Gaiotto:2009we}.
\item[Sec.~\ref{single}:] 
We classify all theories whose gauge group is simple. 
\item[Sec.~\ref{hardpart}\phantom{.5}:] 
We classify the rest, i.e.~theories whose associated graph consists purely of 2-gons and 1-gons.  
\end{itemize}

\subsection{Listing available nodes}\label{nodes}
This is already done by mathematicians in the first half of the last century. $G_i$ is either $\mathsf{A}_r$, $\mathsf{B}_r$, $\mathsf{C}_r$, $\mathsf{D}_r$, or $\mathsf{E}_{6,7,8}$, $\mathsf{F}_4$, $\mathsf{G}_2$.
For our purposes, it turns out to be convenient to describe $\mathsf{B}_r$ and $\mathsf{D}_r$ by just calling them $\SO(n)$. Similarly, we use $\mathsf{A}_{n-1}=\SU(n)$ and $\mathsf{C}_n=\USp(2n)$. 

\subsection{Listing available $n$-gons}\label{ngons}
From \eqref{beta}, available 1-gons are given either by a half-hypermultiplet in a pseudo-real irreducible representation $R$ of $G$ with \begin{equation}
2h^\vee(G) \ge b(R)=c_2(R)
\end{equation} or a full hypermultiplet in a non-pseudo-real representation $R$ of $G$ with \begin{equation}
2h^\vee(G) \ge b(R)=2c_2(R).
\end{equation} We call them usable hypermultiplets.
These are tabulated in Table~\ref{1gon-infinite} and Table~\ref{1gon-isolated}, together with the terminologies for the hypermultiplets which we employ throughout this paper. 
We also list there how many copies of each representations can be added before exceeding $2h^\vee(G)$. 

\begin{table}
\centering
\begin{tabular}{c |c | c | c | c }
half? & Name  & dimension  &  $b$   & How many? \\
\hline
\hline
\multicolumn{5}{c}{$\SU(m)$,  $h^\vee=m$ } \\
\hline
full &   $\fund$ & $m$ & $1$ & $2m$ \\
full, $m\ge 5$ &   $\asym$ & $\frac{m(m-1)}{2}$ & $m-2$ & 3 for $m=5,6$; 2 for $m\geq7$ \\
full &  $\sym$ & $\frac{m(m+1)}{2}$ & $m+2$ & $1$ \\
full, sr &   $\adj$ & $m^2-1$ & $2m$ & $1$ \\
\hline
\multicolumn{5}{c}{$\SO(m)$,  $h^\vee=m-2$} \\
\hline
full, sr &  $\vect$ & $m$ & $2$ & $m-2$  \\
full, sr &  $\adj $ & $m(m-1)/2$ & $2(m-2)$ & 1 \\
\hline
\multicolumn{5}{c}{$\USp(2m)$, $h^\vee=m+1$} \\
\hline
half &  $\vect$ & $m$ & 1/2 &  $4m+4$ \\
full, sr &  $\asym$  & $(2m+1)(m-1)$ & $2(m-1)$ & 3 for $m=2$; 2 for $m=3$; 1 for $m\geq4$ \\
full, sr &  $\adj$ & $m(2m+1)$ & $2(m+1)$ & 1 \\
\hline
\end{tabular}
\caption{List of usable hypermultiplets for single gauge group: infinite series. `sr' stands for `strictly real.'  Dimension is halved for the half-hypermultiplets. 
$\fund$ : fundamental, $\asym$: two-index antisymmetric tensor, $\sym$: two-index symmetric tensor, $\adj$: adjoint, $\vect$: vector. The $\asym$ for $\USp$ is the antisymmetric traceless representation.
We also sometimes call $\vect$ of $\SO$ and $\USp$ as $\fund$, if no confusion arises.  $\asym$ of $\SU(4)$ is treated as an isolated case (see Table~\ref{1gon-isolated}) because it is strictly real.  \label{1gon-infinite}}
\end{table}

\begin{table}
\centering
\begin{tabular}{c| c  | c | c | c |c }
group & half? & Name & Dimension  &  $b$   & How many?  \\
\hline
\hline
$\SU(4)$ & full, sr &  $\asym$ & 6 & 2 &  4 \\
$\SU(6)$ & half &  $\asymT$ & 10 & 3 &  4 \\
$\SU(7)$ & full &   $\asymT$ &35 & 10 &  1 \\
$\SU(8)$ & full &   $\asymT$ &56 & 15 &  1 \\
\hline
$\SO(7)$ & full, sr  &  $\spinor$ & $8$ & $2$ &  5 \\
$\SO(8)$ & full, sr  &  $ \spinor$ & $8$ & $2$ &  6 \\
$\SO(8)$ & full, sr  & $\conj$ & $8$ & $2$ &  6 \\
$\SO(9)$ & full, sr  &  $ \spinor$ & $16$ & $4$ &   3\\
$\SO(10)$ & full    &    $\spinor$ & $16$ & $4$ & 4  \\
$\SO(11)$ & half   &   $\spinor$ & $16$ & $4$ &  4 \\
$\SO(12)$ & half   &   $\spinor$ & $16$ & $4$ &  5 \\
$\SO(12)$ & half    &  $\conj$ & $16$ & $4$ &  5 \\
$\SO(13)$ & half   &   $\spinor$ & $32$ & $8$ &  2 \\
$\SO(14)$ & full  &  $\spinor$ & $64$ & $16$ &  1 \\
\hline
$\USp(4)$  & half &  $\mathbf{16}$  & 8 & 6 &  1 \\
$\USp(6)$  & half &  $\asymT$  & 7 & 5/2 &  3 \\
$\USp(8)$  & half &  $\asymT$ & 24 & 7 &  1 \\
\hline
$\mathsf{E}_6$ & full   &$\mathbf{27}$& $27$ & $6$ & 4 \\
$\mathsf{E}_6$ & full, sr  &$\adj$& $78$ & $24$ & 1 \\
$\mathsf{E}_7$ & half   &$\mathbf{56}$& $28$ & $6$ & 6 \\
$\mathsf{E}_7$ & full, sr  &$\adj$& $133$ & $36$ & 1 \\
$\mathsf{E}_8$ & full, sr  &$\adj$& $248$ & $60$ & 1 \\
$\mathsf{F}_4$ & full, sr  &$\mathbf{26}$& $26$ & $6$ & 3 \\
$\mathsf{F}_4$ & full, sr  &$\adj$& $52$ & $18$ & 1 \\
$\mathsf{G}_2$ & full, sr &$\mathbf{7}$& $7$ & $2$ & 4 \\
$\mathsf{G}_2$ & full, sr  &$\adj$& $14$ & $8$ & 2 \\
\end{tabular}
\caption{List of usable hypermultiplets for single gauge group: isolated ones.
Dimension is halved for the half-hypermultiplets. 
$\asymT$: three-index antisymmetric traceless,
$\spinor$: spinor representation,
$\conj$: another spinor representation,
$\mathbf{16}$, $\mathbf{27}$, $\mathbf{56}$, $\mathbf{26}$, $\mathbf{7}$: representations of the said dimension of $\USp(4)$, $\mathsf{E}_{6}$, $\mathsf{E}_{7}$, $\mathsf{F}_4$ and $\mathsf{G}_2$ respectively.
The $\asymT$ for $\USp$ is the  three-index antisymmetric traceless representation.
\label{1gon-isolated}}
\end{table}

Available 2-gons can be found by noting e.g.~that a full hypermultiplet in $R_1\otimes R_2$  of the group $G_1\times G_2$ contributes to the one-loop beta function of $G_1$ as $\dim R_2$ copies of full hypermultiplets in $R_1$, and looking up the Tables~\ref{1gon-infinite} and \ref{1gon-isolated}.
The result is given in Table~\ref{2-gon}. We note that this table was already compiled thirty years ago \cite{Jiang:1984wr,Jiang:1984nb}.

\begin{table}
\centering
\begin{tabular}{c|r@{$\times$}l|c|c@{$\otimes$}c|cc}
&\multicolumn{2}{c|}{Groups} & half? & \multicolumn{2}{c|}{Name} & condition\\
\hline
\hline
$\checkmark$ &$\SU(n)$ & $\SU(m)$ & full & $\fund$ & $\fund$ & $1/2\le n/m\le 2$\\
$\checkmark$ &$\SO(n)$ & $\USp(2m)$ &  half & $\fund$ & $\fund$ & $n-4\leq 4m $ \& $2m\leq 2n-4$ &\\
\hline
$\circ$ &$\SU(m)$ & $\SO(n)$ & full &$ \fund$ & $\fund$ & $\frac{n}{2}\leq m\leq n-2$\\
$\circ$&$\SU(m)$ & $\USp(2n)$ & full & $\fund$ & $\fund$ & $n\leq m\leq 2(n+1)$ &\\
$\circ$&$\USp(2m)$ & $\USp(2n)$ & full & $\fund$ & $\fund$ & $m-1\leq n\leq m+1$ &\\
\hline
$\bullet$&$\SO(7)$ & $\SU(n)$ & full & $\spinor$ & $\fund$ & $n=4,5$ &\\
$\bullet$&$\SO(7)$ & $\USp(2m)$ & half & $\spinor$ & $\fund$ &$m=1,2,3,4,5$ &\\
$\bullet$&$\mathsf{G}_2$ & $\USp($2m$)$ & half & $\mathbf{7}$ & $\fund$ & $1\leq m\leq4$\\
\hline
$\star$&$\SU(3)$ & $\SU(2)$ & half & $\adj$ & $\fund$ &\\
$\star$ &$\SO(9)$ & $\USp(6)$ & half & $\spinor$ & $\fund$ & &\\
$\star$&$\mathsf{G}_2$ & $\SU(4)$ & full & $\mathbf{7}$ & $\fund$ &
\end{tabular}
\caption{List of usable hypers for $G_1\times G_2$. Two spinor representations of $\SO(8)$ are not listed separately from the vector representation; the five-dimensional representation of $\USp(4)$ is listed as the vector representation of $\SO(5)$; the fundamental of $\SU(2)$ is sometimes listed as the fundamental of $\USp(2)$; the adjoint of $\SU(2)$ is sometimes listed as the vector of $\SO(3)$; the antisymmetric of $\SU(4)$ is regarded as a vector of $\SO(6)$. 
Those marked with $\checkmark$ are the most basic ingredients.
Those marked with $\circ$ arise sporadically.
Those marked with $\bullet$ arise as part of the decoration.
Those marked with $\star$ can only appear in isolation.
 \label{2-gon}}
\end{table}

Available 3-gons can be similarly found based on available 2-gons. We find that only the half-hypermultiplet in $\fund\otimes\fund\otimes\fund$ of  $\SU(2)^3$, or  the half-hypermultiplet in   $\fund\otimes\fund\otimes\fund$ of  $\SU(2)^2\times \USp(4)$ is allowed. They are given in Table~\ref{3-gon}.
The latter, as far as the numerical properties entering in the classification is concerned, can be regarded as a 2-gon between two nodes $\SO(4)$ and $\USp(4)$ corresponding to a half-hypermultiplet in $\vect\otimes \fund$. 
Here we allow $\SO(4)$ to correspond to a node although it is not simple. 
In the following, we regard that there is only one genuine 3-gon that is the half-hypermultiplet in the tri-fundamental representation of $\SU(2)$.  There are no available 4-gons or higher. 

\begin{table}
\centering
\begin{tabular}{l|c|c@{$\otimes$}c@{$\otimes$}c}
\multicolumn{1}{c|}{Groups} & half? & \multicolumn{3}{c}{Name} \\
\hline
\hline
$\SU(2)\times \SU(2)\times\SU(2)$ & half & $\fund$ & $\fund$ & $\fund$ \\
$\SU(2)\times \SU(2)\times\USp(4)$ & half & $\fund$ & $\fund$ & $\fund$ \\
\end{tabular}
\caption{List of usable hypers for $G_1\times G_2\times G_3$. We will regard the latter as $\half\vect\otimes\fund$ of $\SO(4)\times \USp(4)$ by allowing a node representing $\SO(4)$.  \label{3-gon}}
\end{table}

\subsection{Reduction to the classification of conformal theories}\label{reduction}
Our objective is to classify all the ways to combine nodes and $n$-gons so that the one-loop beta function coefficient \eqref{beta} at each node is non-negative (corresponding to the UV-completeness) and integral (corresponding to the freedom from Witten's global anomaly). 
In the following, we identify $n$-gons and corresponding hypermultiplets. 
To reduce the notation, we abbreviate the phrase `$n$ full hypermultiplets in the representation $R$' as $n$ $R$ and `$n$ half-hypermultiplets in the representation $R$' as\footnote{In some culture $n\tfrac12$ means $n$ and a half. We do not mean that here.}  $n$ $\half R$. We do not add `s' denoting the plurals to the name of the representations. So two full hypermultiplets in the adjoint representation is $2\adj$.

The classification process can be simplified by utilizing the following observation.
First, for each $G_i$ we pick a hypermultiplet with the smallest possible $b_i(R_f)$. Explicitly, for $G=\SU(N)$ and $N\ge 3$ it is $\fund$; for $G=\SO(N)$ and $N\ge 7$ it is  $\vect$; for $G=\USp(2N)$ it is $\half\fund$; for exceptional $G$ it can be easily chosen from the Table~\ref{1gon-isolated}.  We call this a \emph{fundamental} hypermultiplet for any $G$, by a slight abuse of terminology.
Then we note that $2h^\vee(G_i)$ and any other $b_i(R')$ is an integral multiple of the contribution $b_i(R_f)$ of the fundamental hypermultiplet. This means that, given a UV-complete $\cN{=}2$ theory and its associated graph, we can add an appropriate number of fundamental hypermultiplets for every node $G_i$ to have a (possibly different) UV-complete $\cN{=}2$ theory so that at every node $G_i$ the one-loop beta function coefficient \eqref{beta} is zero. Then the theory is conformal.\footnote{One way to show this \cite{WestTextbook} is to map the beta function to the R-current anomaly, which is one-loop exact non-perturbatively.  Another is to use the holomorphic renormalization scheme in the $\cN{=}1$ formulation. Then the gauge kinetic term is one-loop exact, and the vectormultiplet wavefunction renormalization is related to it by the extended SUSY. The hypermultiplet is not renormalized \cite{Argyres:1996eh}.}

This argument guarantees that any non-conformal UV-complete theory is obtained by starting from a conformal theory and removing a number of fundamental hypermultiplets from a number of nodes. This removal is at this stage a formal process, but it can be realized physically in most of the cases:
	\begin{itemize}
	\item When the fundamental hypermultiplet of a simple group $G$ is a full hypermultiplet, this can be done by turning on the mass term and making it infinity large.
	\item The fundamental hypermultiplet of a simple group $G$ is a half-hypermultiplet only when $G=\USp(2N)$ or $G=\mathsf{E}_7$. In the former case, due to the condition of the freedom from Witten's global anomaly conditions, we need to remove an even number (say $2n$) of half-hypermultiplets in $\fund$, which can be done by turning on  $n$ mass terms and sending them to infinity. 
	\item When $G=\mathsf{E}_7$ this cannot be done, when we want to remove odd number of half-hypermultiplets in $\mathbf{56}$. So this removal process remains a formal operation. 
	\end{itemize}
When the removal can be done physically, the Seiberg-Witten solution of the theory with less fundamental hypermultiplets can be obtained by taking the infinite mass limit of the corresponding conformal theory. 
The Seiberg-Witten solutions to arbitrary $\mathsf{E}_7$ theories are already known as we review in Sec.~\ref{exceptional}. Therefore, in order to know the Seiberg-Witten solutions to all $\cN{=}2$ gauge theories, it suffices to know solutions to the mass-deformed conformal theories. 

\subsection{Classification of theories with 3-gons}\label{gaiotto}
Recall that there are only two types of usable hypermultiplets charged under three gauge groups, listed in Table~\ref{3-gon}. 
As we stated above, we regard  $\half\fund\otimes\fund\otimes\fund$ of $\SU(2)\times \SU(2)\times \USp(4)$ as  $\half\vect\otimes\fund$ of $\SO(4)\times \USp(4)$ in the combinatorial classification process. Therefore the only possible 3-gon is the one for the \emph{trifundamental} of $\SU(2)$, i.e.~the half-hypermultiplet in the fundamental of $\SU(2)^3$.

Now, suppose we have an $\cN{=}2$ conformal gauge theory with at least one trifundamental in it. 
Pick one of the three $\SU(2)$ node attached to this trifundamental. 
Going over Tables~\ref{2-gon} and \ref{3-gon}, we find that the local structure at the $\SU(2)$ node can be  \begin{itemize}
\item either two trifundamentals attached to it, 
\item or one trifundamental and a 2-gon representing  $\half\fund\otimes\vect$  of $\SU(2)\times \SO(3)$ and a 1-gon representing $\half\fund$ of  $\SU(2)$,
\item or one trifundamental and a 2-gon representing   $\fund\otimes\fund$ of $\SU(2)\times \SU(2)$,
\item or one trifundamental and 1-gons representing  $2\fund$ of $\SU(2)$.
\end{itemize}

We assumed that the associated graph of the theory is connected. Then it automatically follows that every gauge group is $\SU(2)\simeq \SO(3)$ from the analysis above. We find that this is exactly the class of gauge theories found and studied by Gaiotto in his seminal paper \cite{Gaiotto:2009we}. 

\paragraph{Status of the Seiberg-Witten solutions}
The curve for this class of theories is mostly known \cite{Gaiotto:2009we}, except for the following two cases.
One is the theory with the gauge group $\SU(2)^3$ and a full hypermultiplet in the trifundamental representation. 
This allows a mass term for the trifundamental representation. The curve is known if the mass is zero, but it is not known how to obtain the curve for the massive theory. 
Another is the S-dual of this theory, namely the theory with the gauge group $\SO(3)_1\times \SU(2)_0\times \SO(3)_2$, with a bifundamental between $\SO(3)_{1,2}\times \SU(2)_0$  and one additional full fundamental hyper for $\SU(2)_0$.
Again, the curve is known  when the mass term for the full hyper is zero, but otherwise it is unknown. 

Even for the known cases, the curves have  not been derived from the instanton counting, although there have been some attempts toward this direction \cite{Hollands:2010xa,Hollands:2011zc}	.

\subsection{Classification of theories with simple gauge groups}\label{single}
Next, let us classify $\cN{=}2$ conformal theories with only 1-gons. 
Given such a theory,  the gauge group is necessarily simple.  
We first study  theories with adjoint hypermultiplet, and then study other cases one by one for each gauge group. 

\subsubsection{$\cN{=}2^*$ theories}
When the gauge group is $G$ and the theory has a full hypermultiplet in its adjoint representation $\adj$, 
the theory is the $\cN{=}4$  supersymmetric theory with gauge group $G$ when the hypermultiplet is massless,
and is called the $\cN{=}2^*$ theory when the hypermultiplet is massive. 

\paragraph{Status of the Seiberg-Witten solutions}
For $G=\SU(N)$ it was found in Donagi and Witten \cite{Donagi:1995cf}. For the general case, the solution was found in \cite{D'Hoker:1998yi}, but as also discussed in the abstract there, the case $G=\mathsf{G}_2$ does not seem to be completely settled.  

\subsubsection{Theories with other hypermultiplets: exceptional gauge groups}\label{exceptional}
There is no theory with $G=\mathsf{E}_8$. 
For $G=\mathsf{E}_7$, we have six  $\half\mathbf{56}$. 
For $G=\mathsf{E}_6$, we have four  $\mathbf{27}$. 
For $G=\mathsf{F}_4$, we have three   $\mathbf{26}$. 
For $G=\mathsf{G}_2$, we have four  $\mathbf{7}$. 

\paragraph{Status of the Seiberg-Witten solutions}
The Seiberg-Witten geometry with full mass parameters for $G=\mathsf{E}_{6,7}$ was first constructed in \cite{Terashima:1998iz,Hashiba:1999ss} and reviewed in \cite{Tachikawa:2011yr}. The Seiberg-Witten geometry for the theories with $G=\mathsf{F}_4$ was found using the class S technique in \cite{Chacaltana:2014jba}, where the class S derivation for $G=\mathsf{E}_{6}$ was also given.
A class S derivation for $G=\mathsf{E}_7$ was in turn given in \cite{Chacaltana:2017boe}.
The Seiberg-Witten solution for massive $G=\mathsf{G}_2$ theory was found in \cite{Tachikawa:2010vg}. Again this uses the class S technique. 

\subsubsection{Theories with other hypermultiplets: $G=\USp(2n)$} \label{Antonio}
The one-loop beta function is zero in two cases for general $n$: \begin{itemize}
\item $4n+4$ of  $\half\fund$, or
\item  one $\asym$ plus $4\fund$. 
\end{itemize}
There are seven isolated cases: \begin{itemize}
\item $\USp(8)$ theory with one $\half\asymT$ and $3\fund$.
\item $\USp(6)$ theory with $N_a$ $\half\asymT$  plus $N_f$  of  $\half\fund$, for $(N_a,N_f)=(1,11)$, $(2,6)$ and $(3,1)$.
\item $\USp(6)$ theory with one $\half\asymT$ plus one $\asym$   plus three $\half\fund$.
\item $\USp(6)$ theory with two $\asym$.
\item $\USp(4)$ theory with two $\asym$  plus two $\fund$.
\item $\USp(4)$ theory with three $\asym$ -- This is $\SO(5)$ with three $\vect$.
\item $\USp(4)$ theory with one $\half\mathbf{16}$.
\end{itemize}

\paragraph{Status of the Seiberg-Witten solutions}
The Seiberg-Witten solution to $\USp(2n)$ with $4n+4$ half-hypermultiplets in the fundamental, with arbitrary mass parameters, was found in \cite{Argyres:1995fw}. The Seiberg-Witten solution to $\USp(2n)$ with one full massless antisymmetric and $4$ full fundamental hypermultiplets, with arbitrary masses to the fundamental, goes back to  \cite{Douglas:1996js}. The solution to this theory with massive antisymmetric was better understood as a degenerate limit of the linear quiver of the form $\USp(2n)$-$\SU(2n)$-$\cdots$-$\SU(2n)$ terminated with an antisymmetric, which will be described in Sec.~\ref{tree-with-a-long-trunk}. 

As for the isolated cases, 
the $\USp(8)$ case is not known.
Among the isolated $\USp(6)$ cases,
the one  with 11 $\half\fund$ plus $\half\asymT$ and
the one with 3 $\fund$ plus massless $\asymT$ were worked out using class S theory in \cite{Chacaltana:2013oka},
but others are not explicitly known.
However, it would not be impossible to derive the missing $\USp(6)$ cases by combining these class S results and the technique of \cite{Tachikawa:2011yr}.
Finally, first two of the isolated $\USp(4)$ cases can be better thought of as belonging to the family of $\USp(4)$ with $N_a$ $\asym$ and $N_f$ $\half\fund$, where $(N_a,N_f)=(3,0)$, $(2,2)$, $(1,4)$ and $(0,6)$. The last isolated $\USp(4)$ case does not have a known solution.
A uniform class S construction of this family was given in \cite{Chacaltana:2012ch}.

The Seiberg-Witten solution to $\USp(2n)$ with $4n+4$ half-hypermultiplets in the fundamental, with arbitrary mass parameters, was derived by instanton counting in \cite{Shadchin:2004yx}.

\subsubsection{Theories with other hypermultiplets: $G=\SO(n)$}
For general $n$, the one-loop beta function is zero only when we have $n-2$ full hypermultiplets in the vector representation. For sufficiently small $n$, there are a few isolated cases with hypermultiplets in the spinor representation. Up to the identification with the action of the outer automorphism, they are: \begin{itemize}
\item $\SO(14)$ with one $\spinor$ and four $\vect$.
\item $\SO(13)$ with $N_s$ $\half\spinor$  and $N_f$ $\vect$, for $(N_s,N_f)=(2,3)$ and $(N_s,N_f)=(1,7)$.
\item $\SO(12)$ with $N_s$ $\half\spinor$,
$N_c$ $\half\conj$,  and $N_f$ $\vect$, for $(N_s,N_c,N_f)=(5,0,0)$, $(4,1,0)$, $(3,2,0)$; $(4,0,2)$, $(3,1,2)$, $(2,2,2)$; $(3,0,4)$, $(2,1,4)$; $(2,0,6)$, $(1,1,6)$, and $(1,0,8)$. 
\item $\SO(11)$ with $N_s$ $\half\spinor$ and $N_f$ $\vect$, for $(N_s,N_f)=(4,1)$, $(3,3)$, $(2,5)$, $(1,7)$.
\item $\SO(10)$ with $N_s$ $\spinor$ and $N_f$ $\vect$, for $(N_s,N_f)=(4,0)$, $(3,2)$, $(2,4)$, $(1,6)$.
\item $\SO(9)$ with $N_s$ $\spinor$ and $N_f$ $\vect$, for $(N_s,N_f)=(3,1)$, $(2,3)$, $(1,5)$.
\item $\SO(8)$ with $N_s$ $\spinor$,
$N_c$ $\conj$, and  $N_f$ $\vect$, for $(N_s,N_c,N_f)=(5,1,0)$, $(4,2,0)$, $(4,1,1)$, $(3,3,0)$, $(3,2,1)$, $(2,2,2)$. 
\item $\SO(7)$ with $N_s$ $\spinor$ 
and $N_f$ $\vect$, for $(N_s,N_f)=(5,0)$, $(4,1)$, $(3,2)$, $(2,3)$, $(1,4)$.
\end{itemize}

\paragraph{Status of the Seiberg-Witten solutions}

For the general case with $n-2$ hypermultiplets, the solution for arbitrary masses was found in  \cite{Argyres:1995fw,Hanany:1995fu}.
For $\SO(7)$ with spinors,  the solutions to all possibilities were found in \cite{Chacaltana:2011ze,Chacaltana:2013oka}.
For $\SO(8)$, $\SO(10)$ and $\SO(12)$ with spinors, the solutions to all possibilities with arbitrary mass parameters were found in  \cite{Terashima:1998fx} and reviewed in \cite{Tachikawa:2011yr}.
The second construction for $\SO(8)$ theories with spinors   was given in \cite{Chacaltana:2011ze,Chacaltana:2013oka}. 
For $\SO(9)$, $\SO(11)$ with spinors, the solution was given in \cite{Chacaltana:2014ica}.
For $\SO(13)$, the solution for $(N_s,N_f)=(1,7)$ was given in \cite{Chacaltana:2014ica} using the class S technique. 
The solution for $(N_s,N_f)=(2,3)$ was not available using the class S technique, but it would not be impossible to apply the method of \cite{Tachikawa:2011yr} to the $(N_s,N_f)=(1,7)$ solution to find the solution for this one.
Finally, no solution for the $\SO(14)$ with spinors is known.

\subsubsection{Theories with other hypermultiplets: $G=\SU(n)$}

The one-loop beta function is zero in five cases for general $n$: \begin{itemize}
\item $2n$ of $\fund$.
\item one $\asym$ plus $n+2$ $\fund$. 
\item two $\asym$, plus $4$ $\fund$. 
\item one  $\sym$, plus $n-2$ $\fund$. 
\item one  $\sym$ and one $\asym$.
\end{itemize}
There are a few isolated cases for $n=4,5,6,7,8$: \begin{itemize}
\item $\SU(8)$ theory with one $\asymT$ ,
and one $\fund$.
\item $\SU(7)$ theory with one $\asymT$ and four $\fund$.
\item $\SU(6)$ theory with $N_3$ $\half\asymT$, $N_2$ $\asym$ and $N_f$  $\fund$,  for $(N_3,N_2,N_f)=(4,0,0)$, $(3,0,3)$, $(2,0,6)$, $(1,0,9)$, $(2,1,2)$, $(1,1,5)$ and $(1,2,1)$.
\item $\SU(6)$ theory with 3 $\asym$
\item $\SU(5)$ theory with 3 $\asym$ and one $\fund$
\item $\SU(4)$ theory with 3 $\asym$ and two $\fund$
\item $\SU(4)$ theory with 4 $\asym$ -- This is $\SO(6)$ is 4 $\vect$.
\end{itemize}

\paragraph{Status of the Seiberg-Witten solutions}

The Seiberg-Witten curves for $\SU(n)$ with arbitrary number of hypermultiplets in the fundamental representation were determined in \cite{Hanany:1995na,Argyres:1995wt}.
Those  for $\SU(n)$ with  hypermultiplets in the antisymmetric tensor and in the fundamental representation were found and described in detail in \cite{Argyres:2002xc}.
Those  for $\SU(n)$ with  hypermultiplets in the symmetric tensor and in the fundamental representation were found in \cite{Landsteiner:1997vd,Landsteiner:1998pb}.
The curve  for $\SU(n)$ with  one hypermultiplet in the symmetric tensor and another in the antisymmetric tensor representation  was found in \cite{Chacaltana:2012ch}.

For the isolated cases, 
the solutions to $\SU(8)$ or $\SU(7)$  theories with a hypermultiplet in the three-index antisymmetric tensor are not known yet.
The solutions to $\SU(6)$ and $\SU(5)$ theories can be written down using the technique in \cite{Tachikawa:2011yr}.
All `isolated' solutions for $\SU(4)$ are known, by combining \cite{Gaiotto:2009we} and \cite{Tachikawa:2009rb}.

For all cases, the instanton counting described in \cite{Nekrasov:2002qd,Shadchin:2004yx} is in principle possible, and the Seiberg-Witten curves for theories without $\asymT$ and with at most two $\asym$ were indeed derived there.  It would be interesting to obtain the Seiberg-Witten curves for the remaining isolated cases in this way.

\subsubsection{Pure theories}
In our classification method, we reduced the general, non-conformal theories to the classification of the conformal theories.
But before proceeding further, it is not completely useless to discuss the Seiberg-Witten solutions of the pure theory, where we just have a gauge multiplet in a simple group $G$. 

\paragraph{Status of the Seiberg-Witten solutions}
The solution to the $\SU(2)$ case  was found in the original paper \cite{Seiberg:1994rs}.
This was extended to simply-laced gauge groups in \cite{Klemm:1994qs}.
The solution to the $\SU(n)$ case was also discussed in \cite{Argyres:1994xh}.
The solution to the first non-simply-laced case was the one for $\SO(2r+1)$ in \cite{Danielsson:1995is}.
The solution to the pure $\SO(2n)$ theory  was discussed in detail in \cite{Brandhuber:1995zp}.
A general solution applicable to any pure $G$ theory was found in \cite{Martinec:1995by}, and the structure of the Seiberg-Witten curves of this form was studied in detail in \cite{Hollowood:1997pp}.

\section{The classification, part II}\label{hardpart}
In this section we perform the remaining, most significant part of the classification. Namely, we study theories with semisimple gauge groups, which involve 2-gons and 1-gons. 
We go through the following steps: \begin{itemize}
\item[Sec.~\ref{veryrare}:] We first deal with theories with very rare type of 2-gons, marked by $\star$ in Table~\ref{2-gon}. 
\item[Sec.~\ref{overall}:] We then perform the main classification. Given an $\cN{=}2$ theory not taken care of thus far, we take its associated graph, and consider its \emph{main subgraph}, consisting only of the 2-gons corresponding to (full or half) hypermultiplets in $\fund\otimes\fund$ of some $G_1\times G_2$, marked by $\checkmark$ and $\circ$ in Table~\ref{2-gon}.
We assign a \emph{current} for each  edge corresponding to a (full or half) hypermultiplet in $\fund\otimes\fund$ of $G_1\times G_2$ within the main subgraph.  By an easy argument, we find that two or more positive currents never flow in to a node, and with a slightly more argument, we find that the main subgraph is either a single loop or a tree.  Then, we will see that the associated graph always consists of a few types of \emph{trunks} to which a number of \emph{branches} are attached, where  the 2-gons and the 1-gons other than the full or half hypermultiplet in $\fund\otimes\fund$ are treated as various \emph{decorations}.
\item[Sec.~\ref{branches}:] The task is then to enumerate all possible branches, and 
\item[Sec.~\ref{combinations}:] to enumerate all possible ways to combine branches. 
\end{itemize}
After the classificaiton, we describe in Sec.~\ref{statusX} the status of the Seiberg-Witten solutions of these theories.

Before starting the actual work, let us set up some notations we employ in the pictures below. 
We use 
\begin{tikzpicture}
\node[SU] (A) at (0,0) {$m$};
\end{tikzpicture}
 to denote an $\SU(m)$ node, 
\begin{tikzpicture}
\node[Sp] (A) at (0,0) {$2m$} ;
\end{tikzpicture}
to denote a $\USp(2m)$ vertex, and
\begin{tikzpicture}
\node[SO] (A) at (0,0) {$m$} ;
\end{tikzpicture}
to denote an $\SO(m)$ vertex. As $\SU(4)\simeq \SO(6)$ and $\USp(4)\simeq \SO(5)$, we need to keep in mind that sometimes 
\begin{tikzpicture}
\node[SU] (A) at (0,0) {$4$};
\end{tikzpicture} can serve as 
\begin{tikzpicture}
\node[SO] (A) at (0,0) {$6$} ;
\end{tikzpicture}\ , etc. To represent these identification graphically, we sometimes use the symbol 
\begin{tikzpicture}
\node[SU] (A) at (0,0) {$4$};
\node[SO] (B) at (.5,0) {$6$};
\end{tikzpicture}
 for one node representing $\SU(4)\simeq\SO(6)$, and 
\begin{tikzpicture}
\node[SO] (A) at (0,0) {$5$};
\node[Sp] (B) at (.5,0) {$4$};
\end{tikzpicture}
 for one node representing $\SO(5)\simeq\USp(4)$.

We denote 2-gons corresponding to $\half\fund\otimes\fund$ of $\USp(m)\times\SO(n)$ and $\fund\otimes\fund$ of other $G_1\times G_2$ by straight edges between corresponding vertices. For other 2-gons, we add annotations to the edges. 
As for one-gons, we do not draw the 1-gons representing the fundamental hypermultiplets. As discussed in Sec.~\ref{reduction}, they can always be added in appropriate numbers so as to make all the one-loop beta functions to vanish.  The other 1-gons are denoted by connecting e.g.~\begin{tikzpicture}
\node[1gon] (A) at (0,0) {$\asym$} ;
\end{tikzpicture} to a node. In the case of an identified node, the full classification is obtained by allowing the addition of flavors to both sides of an identified node. For instance, the last entry in Sec. \ref{Antonio} $\USp(4)$ with two $\asym$ and two $\fund$ can be written alternatively as
 \begin{tikzpicture}
 \node[1gon] (C) at (-1,0) {$2$} ;
\node[SO] (A) at (0,0) {$5$};
\node[Sp] (B) at (.5,0) {$4$};
\node[1gon] (D) at (1.5,0) {$2$} ;
\draw (C)--(A);
\draw (D)--(B);
\end{tikzpicture}
which represents two $\fund$ for $\USp(4)$ and two $\fund$ for $\SO(5)$.

Before proceeding, we note that some partial classification was done thirty years ago in \cite{Jiang:1984wn,Jiang:1984aa}. 
For theories with gauge group $\prod_i\SU(n_i)$ and bifundamental hypermultiplets,
it has been long known that their classification is equivalent to the classification of Dynkin diagrams \cite{Katz:1997eq}.

\subsection{Three rare cases}\label{veryrare}
We first consider  theories with  2-gons of the type marked by $\star$ in Table~\ref{2-gon}, namely $\half\adj\otimes\fund$ of $\SU(3)\times\SU(2)$,
$\half\spinor\otimes\fund$ of $\SO(9)\times\USp(6)$, and
$\mathbf{7}\otimes\fund$ of $\mathsf{G}_2\times \SU(4)$. 
It is clear that the only theories that involve them are: \begin{itemize}
\item 
\begin{tikzpicture}
\node[SU] (A) at (0,0) {$3$};
\node[SU] (B) at (3,0) {$2$};
\draw (A)-- node[above] {$\half\adj\otimes\fund$} (B);
\end{tikzpicture}\ ,
\item 
\begin{tikzpicture}
\node[SO] (A) at (0,0) {$9$};
\node[Sp] (B) at (3,0) {$6$};
\draw (A)-- node[above] {$\half\spinor\otimes\fund$} (B);
\end{tikzpicture}
with one half-hypermultiplet in the fundamental of $\USp(6)$, and 
\item 
\begin{tikzpicture}
\node (A) at (0,0) {$\mathsf{G}_2$};
\node[SU] (B) at (3,0) {$4$};
\draw (A)-- node[above] {$\mathbf{7}\otimes\fund$} (B);
\end{tikzpicture}
with one fundamental of $\SU(4)$.
\end{itemize}

\subsection{Overall structure of the remaining theories}\label{overall}

Given an $\cN{=}2$ theory not analyzed so far, let us take its associated graph, and consider its \emph{main subgraph}, consisting only of the 2-gons corresponding to (full or half) hypermultiplets in $\fund\otimes\fund$ of some $G_1\times G_2$, marked by $\checkmark$ and $\circ$ in Table~\ref{2-gon}. We call them the bifundamental hypermultiplets.
For each  edge corresponding to a (full or half) hypermultiplet in the bifundamental of $X(m)\times Y(n)$ where $X,Y=\SU,\USp,\SO$, we define the \emph{current} at the edge to be $(m+2)-n$ for $\half\fund\otimes\fund$ of $\USp(m)\times \SO(n)$,
and $m-n$ otherwise.
	We say that the current $|m-n|$ flows from the node $X(m)$ to the node $Y(n)$ when $m>n$ in the second case, and similarly for the former.

\begin{itemize}
\item 
Going over each of the bifundamentals carefully, one finds that
	\begin{itemize}
	\item  when a positive current flows in to a node $X(m)$ through an edge representing a (full or half) bifundamental hypermultiplet $R$, we have \begin{equation}
h^\vee(X(m)) < b_{X(m)}(R),
\end{equation}  
	\item when there is zero current between the nodes  $X(m)$ and $Y(n)$ through an edge representing a (full or half) bifundamental hypermultiplet $R$, we have \begin{equation}
h^\vee(X(m)) \le b_{X(m)}(R).
\end{equation}
	\end{itemize}
Therefore, at a node, 
	\begin{itemize}
	\item there is at most one positive current flowing in,
	\item if there is a positive current flowing in, there cannot be an edge with zero current, 	\item and if two bifundamental edges with zero currents meet at a node $X(m)$, the node $X(m)$ is saturated and there is no other $1$-gons or $2$-gons attached to the node. The possibilities are
		\begin{itemize}
	\item \begin{tikzpicture}
\node[SU] (A) at  (1,0)  {$m$};
\node[SU] (B) at  (2,0)  {$m$};
\node[SU] (C) at  (3,0)  {$m$};
\draw (A) --(B)--(C);
\end{tikzpicture}\ , 
	\item \begin{tikzpicture}
\node[SO] (A) at  (1,0)  {$m$};
\node[Sp] (B) at  (2.5,0)  {$m-2$};
\node[SO] (C) at  (4,0)  {$m$};
\draw (A) --(B)--(C);
\end{tikzpicture}\ , 
\begin{tikzpicture}
\node[Sp] (A) at  (1,0)  {$m-2$};
\node[SO] (B) at  (2.5,0)  {$m$};
\node[Sp] (C) at  (4,0)  {$m-2$};
\draw (A) --(B)--(C);
\end{tikzpicture}\ ,   
	\item \begin{tikzpicture}
\node[SU] (A) at  (1,0)  {$4$};
\node[SU] (B) at  (2.25,0)  {$4$};
\node[SO] (C) at  (2.75,0)  {$6$};
\node[Sp] (D) at  (4,0)  {$4$};
\draw (A) --(B);
\draw (C) --(D);
\end{tikzpicture}  where the overlapping $\SU(4)$ and $\SO(6)$ means that they are identified, 
	\item \begin{tikzpicture}
\node[SU] (A) at  (1,0)  {$2m$};
\node[SU] (B) at  (2.25,0)  {$2m$};
\node[Sp] (C) at  (3.5,0)  {$2m$};
\draw (A) --(B)--(C);
\end{tikzpicture}\ , or
\begin{tikzpicture}
\node[Sp] (A) at  (1,0)  {$2m$};
\node[SU] (B) at  (2.25,0)  {$2m$};
\node[Sp] (C) at  (3.5,0)  {$2m$};
\draw (A) --(B)--(C);
\end{tikzpicture}\ .
	\end{itemize}
	\end{itemize}
\end{itemize}

From these considerations, we understand that any theory has the following structure:
\begin{itemize}
\item If there is a loop of bifundamentals in the main subgraph, the associated graph itself consists only of this single loop, without any addition. Every node is  already conformal, and no other hypermultiplets can be added.  
\item Otherwise, the main subgraph is a tree, which forces the associated graph itself to be a tree. We define the \emph{trunk} to be the part of the main subgraph where the currents are zero, and the \emph{branches} to be the rest of the associated graph.
A branch  can only start at the ends of the trunk. 
In practice, it is more convenient to allow branches where the current can be zero; this facilitates the analysis of what can be put at one end of a long trunk. So we use this modified definition of branches below. 
\item The possible types of trunks can be easily found. It is either
	\begin{itemize}
	\item  just a single node, either $\SU(m)$, $\SO(m)$, $\USp(m)$ or $\mathsf{G}_2$,
	\item  an $\SU(m)$ chain
\begin{tikzpicture}
\node (0) at (0,0) {};
\node[SU] (A) at (1,0) {$m$};
\node[SU] (B) at (2,0) {$m$};
\node[SU] (C) at (3,0) {$m$};
\node[SU] (D) at (4,0) {$m$};
\node (E) at (5,0) {};
\draw[dashed] (0)--(A);
\draw[dashed] (D)--(E);
\draw (A)--(B)--(C)--(D);
\end{tikzpicture}\ ,
	\item  an $\SO(m)$-$\USp(m-2)$ chain
\begin{tikzpicture}
\node (0) at (0,0) {};
\node[SO] (A) at (1,0) {$m$};
\node[Sp] (B) at (2.5,0) {$m-2$};
\node[SO] (C) at (4,0) {$m$};
\node[Sp] (D) at (5.5,0) {$m-2$};
\node (E) at (7,0) {};
\draw[dashed] (0)--(A);
\draw[dashed] (D)--(E);
\draw (A)--(B)--(C)--(D);
\end{tikzpicture}\ ,
	\item a mixture of these two when they involve $\SU(4)$ and $\SO(6)$.
	
	 One example is 
	\begin{tikzpicture}
\node (G) at  (-1,0)  {};
\node[SU] (F) at  (0,0)  {$4$};
\node[SU] (A) at  (1,0)  {$4$};
\node[SU] (B) at  (2,0)  {$4$};
\node[SO] (C) at  (2.5,0)  {$6$};
\node[Sp] (D) at  (3.5,0)  {$4$};
\node[SO] (E) at  (4.5,0)  {$6$};
\node (Q) at  (5.5,0)  {};
\draw[dashed] (G)--(F);
\draw (F)--(A)--(B);
\draw (C)--(D)--(E);
\draw[dashed] (E)--(Q);
\end{tikzpicture}  where the two overlapping nodes mean that they are to be identified.
	\end{itemize}
We treat an end of the $\SU(2m)$ trunk terminated by a $\USp(2m)$ as a branch, as this occurs only rarely.
\end{itemize}

Therefore, our remaining tasks are to classify branches and then to enumerate all possible ways to combine branches. We do this in Sec.~\ref{branches} and Sec.~\ref{combinations} in turn.

\subsection{Branches}\label{branches}
In this section we classify the branches. 
Our convention is to call the node of the branch which is shared with the trunk the zero-th node.
The length of a branch is the length of the longest chain of the edges. 
We also assign a \emph{branch current} to each branch, defined as follows. 
We take the difference between $h^\vee$ of the zero-th node of the branch minus the beta function contribution from the first node, and multiply the difference by $1$, $1$ and $2$ depending on whether the zero-th node is of type $\SU$, $\SO$ and $\USp$,  respectively. 
For the branches starting with $\SU-\SU$, $\USp-\SO$, $\SO-\USp$, $\SU-\USp$  and $\SU-\SO$ bifundamentals, it is  equal to the \emph{current}.
A branch is called \emph{small} when the branch current is non-negative. Those whose branch current is negative are called \emph{large}. 

Now, let us list all possible branches.  Clearly, we only need to find maximal branches, in the sense that no further nodes can be added to the end farthest from the trunk.  Therefore, non-maximal branches are not necessarily listed, to make the presentation more concise.  A branch whose zero-th node represents a group $X$ is called an $X$ branch;
similarly, a branch which starts with groups $X$ and $Y$ is called an $X$-$Y$ branch. 
We start from rare large branches with length 1, and then visit each of $X$ (or $X$-$Y$) branches one by one.

\subsubsection{Large branches with length 1}\label{what?}
First, we list rare examples of large branches with length 1. Going over Table~\ref{1gon-infinite} and Table~\ref{1gon-isolated}, we find that they are
\begin{itemize}
\item
\begin{tikzpicture}
\node[1gon] (A) at (0,0) {$\sym$};
\node[SU] (B) at (-2,0) {$n$};
\draw (A)--(B);
\end{tikzpicture}\ ,
\item
\begin{tikzpicture}
\node[1gon] (A) at (0,0) {$\asymT$};
\node[SU] (B) at (-2,0) {$7$};
\draw (A)--(B);
\end{tikzpicture}\ ,
\item
\begin{tikzpicture}
\node[1gon] (A) at  (2,0)  {$\spinor$};
\node[SO] (B) at  (0,0)  {14};
\draw (A)--(B);
\end{tikzpicture}\ ,
\item
\begin{tikzpicture}
\node[1gon] (A) at  (0,0)  {$\asymT$};
\node[Sp] (B) at  (-2,0)  {8};
\draw (A)--(B);
\end{tikzpicture}\ .
\end{itemize}

\subsubsection{$\SU$-branches (all small)}
\paragraph{Undecorated version}
The basic structure is 
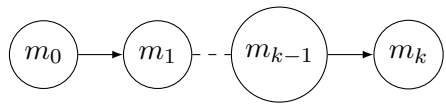
\[
\begin{tikzpicture}
\node[SU] (B) at  (2,0)  {$m_0$};
\node[SU] (C) at  (3.5,0)  {$m_1$};
\node[SU] (D) at  (5.1,0)  {$m_{ k-1}$};
\node[SU] (E) at  (6.8,0)  {$m_k$};
\draw[->] (B)--(C);
\draw[dashed] (C)--(D);
\draw[->] (D)--(E);
\end{tikzpicture} 
\] where a directed edge is a reminder of the flow direction of non-negative current. 
An edge is not directed if the current along the edge is exactly zero or there does not exist a defined notion of current for the two types of vertices that the edge connects. The latter case corresponds to various \emph{decorations} as we will see below.

The current $I_i=m_i-m_{i+1}$ is non-decreasing as we go along the branch. 
The branch current  is, by definition, equal to $I_0$. The branch current is $0 \le I_0 \le m_0/2$.
When the current of the branch is $\le 3$, we can have various \emph{decorations}, which we list below:

\paragraph{Decorations with $\USp$:}
When the branch current is $0$, we can have 
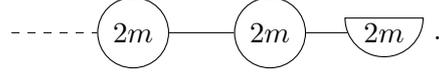
\[
\begin{tikzpicture}
\node[SU] (D) at  (5.1,0)  {$2m$};
\node[SU] (E) at  (6.9,0)  {$2m$};
\node[Sp] (F) at  (8.4,0)  {$2m$};
\draw[dashed] (3.5,0)--(D);
\draw (D)--(E);
\draw (E)--(F);
\end{tikzpicture}\ .
\]

This can be extended in two rare cases: 
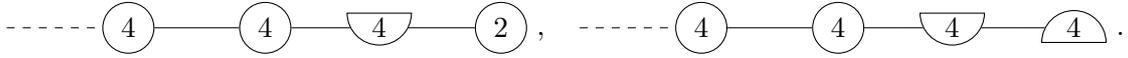
\[
\begin{tikzpicture}
\node[SU] (D) at  (5.1,0)  {$4$};
\node[SU] (E) at  (6.9,0)  {$4$};
\node[Sp] (F) at  (8.4,0)  {$4$};
\node[SU] (G) at  (10,0)  {$2$};
\draw[dashed] (3.5,0)--(D);
\draw (D)--(E);
\draw (E)--(F)--(G);
\end{tikzpicture}\ ,  \quad
\begin{tikzpicture}
\node[SU] (D) at  (5.1,0)  {$4$};
\node[SU] (E) at  (6.9,0)  {$4$};
\node[Sp] (F) at  (8.4,0)  {$4$};
\node[SO] (G) at  (10,0)  {$4$};
\draw[dashed] (3.5,0)--(D);
\draw (D)--(E);
\draw (E)--(F)--(G);
\end{tikzpicture}\ .
\] 

When the branch current is $0,1$, its end can be decorated: 
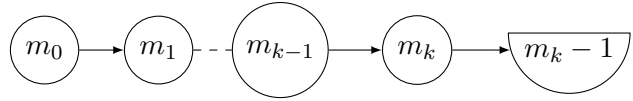
\[
\begin{tikzpicture}
\node[SU] (B) at  (2,0)  {$m_0$};
\node[SU] (C) at  (3.5,0)  {$m_1$};
\node[SU] (D) at  (5.1,0)  {$m_{k-1}$};
\node[SU] (E) at  (6.9,0)  {$m_k$};
\node[Sp] (F) at  (8.9,0)  {$m_k-1$};
\draw[->] (B)--(C);
\draw[dashed] (C)--(D);
\draw[->] (D)--(E);
\draw[->] (E)--(F);
\end{tikzpicture}
\] where $m_k$ needs to be odd, and $k$ can be zero.

When the branch current  is $0,1,2$, its end can be decorated: 
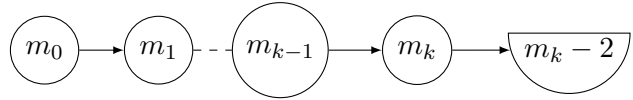
\[
\begin{tikzpicture}
\node[SU] (B) at  (2,0)  {$m_0$};
\node[SU] (C) at  (3.5,0)  {$m_1$};
\node[SU] (D) at  (5.1,0)  {$m_{k-1}$};
\node[SU] (E) at  (6.9,0)  {$m_k$};
\node[Sp] (F) at  (8.9,0)  {$m_k-2$};
\draw[->] (B)--(C);
\draw[dashed] (C)--(D);
\draw[->] (D)--(E);
\draw[->] (E)--(F);
\end{tikzpicture} 
\]
where $m_k$ needs to be even, and $k$ can be zero.

\paragraph{Decorations with some 1-gon:}
If the branch current is $0,1,2$, we can have: 
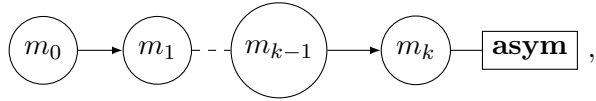
\[
\begin{tikzpicture}
\node[SU] (B) at  (2,0)  {$m_0$};
\node[SU] (C) at  (3.5,0)  {$m_1$};
\node[SU] (D) at  (5.1,0)  {$m_{k-1}$};
\node[SU] (E) at  (6.9,0)  {$m_k$};
\node[1gon] (F) at  (8.4,0)  {$\asym$};
\draw[->] (B)--(C);
\draw[dashed] (C)--(D);
\draw[->] (D)--(E);
\draw (E)--(F);
\end{tikzpicture}\ ,
\] 
and when the current of the branch is $0,1,2,3$, its end can be decorated: 
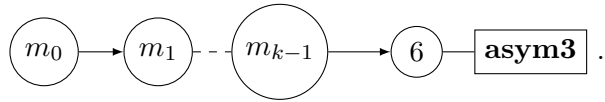
\[
\begin{tikzpicture}
\node[SU] (B) at  (2,0)  {$m_0$};
\node[SU] (C) at  (3.5,0)  {$m_1$};
\node[SU] (D) at  (5.1,0)  {$m_{k-1}$};
\node[SU] (E) at  (6.9,0)  {$6$};
\node[1gon] (F) at  (8.4,0)  {$\asymT$};
\draw[->] (B)--(C);
\draw[dashed] (C)--(D);
\draw[->] (D)--(E);
\draw (E)--(F);
\end{tikzpicture}\ .
\]

\paragraph{Further branching:}
The only possibility is 
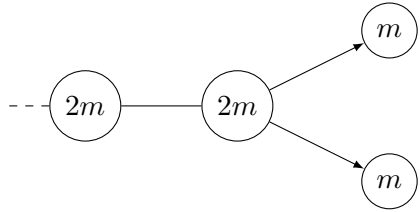
\[
\begin{tikzpicture}
\node[SU] (0) at  (-2,0)  {$2m$};
\node[SU] (A) at  (0,0)  {$2m$};
\node[SU] (B) at  (2,1)  {$m$};
\node[SU] (C) at  (2,-1)  {$m$};
\draw[dashed] (-3,0)--(0);
\draw[-] (0)--(A);
\draw[->] (A)--(B);
\draw[->] (A)--(C);
\end{tikzpicture}
\] with zero branch current.
When $m=2$, one or two $\SU(2)$ can be replaced by $\asym$ of $\SU(4)$. When $m=3$, one or two $\SU(3)$ can be replaced by $\asymT$ of $\SU(6)$.

\paragraph{Branches involving identifications:}
We have the following two possibilities: 
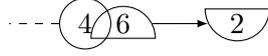
\[
\begin{tikzpicture}
\node[SU] (A) at  (0,0)  {4};
\node[SO] (B) at  (.5,0)  {6};
\node[Sp] (C) at  (2,0)  {2};
\draw[->] (B)--(C);
\draw[dashed] (A)--(-1,0);
\end{tikzpicture}
\] for an $\SU$ branch with branch current $\leq2$, and
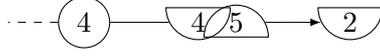
\[
\begin{tikzpicture}
\node[SU] (Z) at  (-1.5,0)  {4};
\node[Sp] (A) at  (0,0)  {4};
\node[SO] (B) at  (.5,0)  {5};
\node[Sp] (C) at  (2,0)  {2};
\draw (Z)--(A);
\draw[->] (B)--(C);
\draw[dashed] (Z)--(-2.5,0);
\end{tikzpicture}
\] for an $\SU(4)-\SU(4)$ branch with branch current $=0$.
Here, as before, the overlapping nodes mean that they should be identified. 

\subsubsection{$\USp$-$\SO$ and $\SO$-$\USp$ branches (all small)}
\paragraph{Undecorated branches}
The basic cases are 
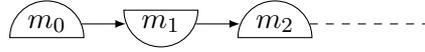
\[
\begin{tikzpicture}
\node[SO] (B) at  (2,0)  {$m_0$};
\node[Sp] (C) at  (3.5,0)  {$m_1$};
\node[SO] (D) at  (5,0)  {$m_2$};
\draw[->] (B)--(C);
\draw[->] (C)--(D);
\draw[dashed] (D)--(7,0);
\end{tikzpicture} 
\] and \[
\begin{tikzpicture}
\node[Sp] (B) at  (2,0)  {$m_0$};
\node[SO] (C) at  (3.5,0)  {$m_1$};
\node[Sp] (D) at  (5,0)  {$m_2$};
\draw[->] (B)--(C);
\draw[->] (C)--(D);
\draw[dashed] (D)--(7,0);
\end{tikzpicture}\ .
\] 
The current $I_i=\tilde m_i - \tilde m_{i+1}$ where $\tilde m=m$  for $\SO$, $\tilde m=m+2$ for $\USp$,
is non-decreasing as we go along the branch.
The branch current is, by definition, equal to $I_0$.
Starting with $\SO(m)$ or $\USp(m)$, the branch current is at most $m/2$. 

When the current in the branch is $\le 6$, we can have various decorations. 

\paragraph{Decorations with one $\SU$ at the end:}
If the branch current $=0$, we can have: 
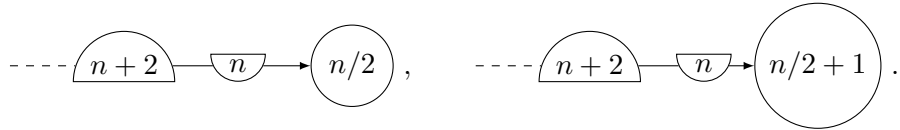
\[
\begin{tikzpicture}
\node[SO] (A) at (0,0) {$n+2$};
\node[Sp] (B) at (1.5,0) {$n$};
\node[SU] (C) at (3,0) {$n/2$};
\draw[dashed] (-1.5,0)--(A);
\draw (A)--(B);
\draw[->] (B)--(C);
\end{tikzpicture}\ , \qquad
\begin{tikzpicture}
\node[SO] (A) at (0,0) {$n+2$};
\node[Sp] (B) at (1.5,0) {$n$};
\node[SU] (C) at (3,0) {$n/2+1$};
\draw[dashed] (-1.5,0) -- (A);
\draw (A)--(B);
\draw[->] (B)--(C);
\end{tikzpicture}\ .
\]

When $n=4,6$, they can be further extended: 
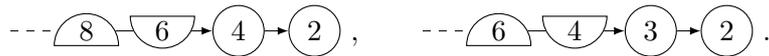
\[
\begin{tikzpicture}
\node[SO] (A) at (0,0) {$8$};
\node[Sp] (B) at (1,0) {$6$};
\node[SU] (C) at (2,0) {$4$};
\node[SU] (D) at (3,0) {$2$};
\draw[dashed] (-1,0)--(A);
\draw (A)--(B);
\draw[->] (B)--(C);
\draw[->] (C)--(D);
\end{tikzpicture}\ , \qquad 
\begin{tikzpicture}
\node[SO] (A) at (0,0) {$6$};
\node[Sp] (B) at (1,0) {$4$};
\node[SU] (C) at (2,0) {$3$};
\node[SU] (D) at (3,0) {$2$};
\draw[dashed] (-1,0)--(A);
\draw (A)--(B);
\draw[->] (B)--(C);
\draw[->] (C)--(D);
\end{tikzpicture}\ .
\]

If the branch current $\leq1$, we can have: 
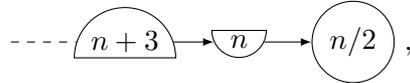
\[
\begin{tikzpicture}
\node[SO] (A) at (0,0) {$n+3$};
\node[Sp] (B) at (1.5,0) {$n$};
\node[SU] (C) at (3,0) {$n/2$};
\draw[dashed] (-1.5,0)--(A);
\draw[->] (A)--(B);
\draw[->] (B)--(C);
\end{tikzpicture}\ ,
\]
and if the branch current $\leq2$, we can have: 
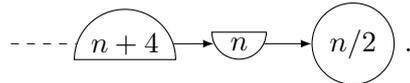
\[
\begin{tikzpicture}
\node[SO] (A) at (0,0) {$n+4$};
\node[Sp] (B) at (1.5,0) {$n$};
\node[SU] (C) at (3,0) {$n/2$};
\draw[dashed] (-1.5,0)--(A);
\draw[->] (A)--(B);
\draw[->] (B)--(C);
\end{tikzpicture}\ .
\]

\paragraph{Decorations with other 1-gons and 2-gons:}
In this paragraph, the leftmost node is given as an illustration. The 1-gons corresponding to  spinors of $\SO(7)$ or $\SO(8)$ are not listed; they can be freely replaced with the vectors  and vice-versa.

If the branch current $=0$, we can have the structure 
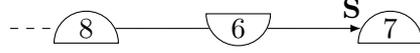
\[
\begin{tikzpicture}
\node[SO] (A) at (0,0) {$8$};
\node[Sp] (B) at (2,0) {$6$};
\node[SO] (C) at (4,0) {$7$};
\draw[dashed] (-1,0)--(A);
\draw (A)--(B);
\draw[->] (B)-- node[above, pos=0.9]{$\spinor$} (C);
\end{tikzpicture}
\]
where $\spinor$ at the end of $\USp(6)-\SO(7)$ edge signifies that the edge stands for $\half\fund\otimes\spinor$ of $\USp(6)\times\SO(7)$.  
Maximal branches containing this structure are 

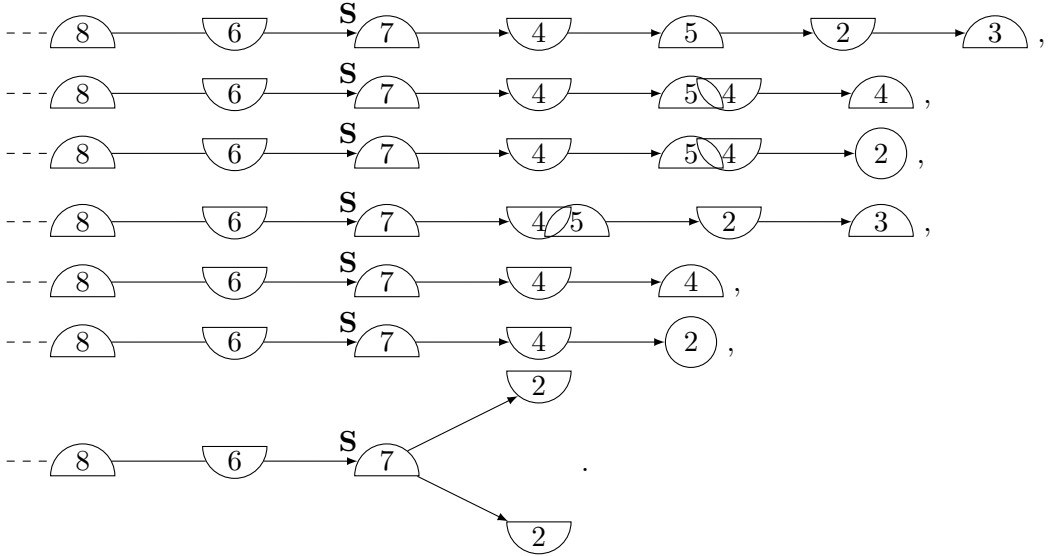
\begin{tikzpicture}
\node[SO] (A) at (0,0) {$8$};
\node[Sp] (B) at (2,0) {$6$};
\node[SO] (C) at (4,0) {$7$};
\node[Sp] (D) at (6,0) {$4$};
\node[SO] (E) at (8,0) {$5$};
\node[Sp] (F) at (10,0) {$2$};
\node[SO] (G) at (12,0) {$3$};
\draw[dashed] (-1,0)--(A);
\draw (A)--(B);
\draw[->] (B)-- node[above, pos=0.9]{$\spinor$} (C);
\draw[->] (C)--(D);
\draw[->] (D)--(E);
\draw[->] (E)--(F);
\draw[->] (F)--(G);
\end{tikzpicture}\ ,

\begin{tikzpicture}
\node[SO] (A) at (0,0) {$8$};
\node[Sp] (B) at (2,0) {$6$};
\node[SO] (C) at (4,0) {$7$};
\node[Sp] (D) at (6,0) {$4$};
\node[SO] (E) at (8,0) {$5$};
\node[Sp] (F) at (8.5,0) {$4$};
\node[SO] (G) at (10.5,0) {$4$};
\draw[dashed] (-1,0)--(A);
\draw (A)--(B);
\draw[->] (B)-- node[above, pos=0.9]{$\spinor$} (C);
\draw[->] (C)--(D);
\draw[->] (D)--(E);
\draw[->] (F)--(G);
\end{tikzpicture}\ ,

\begin{tikzpicture}
\node[SO] (A) at (0,0) {$8$};
\node[Sp] (B) at (2,0) {$6$};
\node[SO] (C) at (4,0) {$7$};
\node[Sp] (D) at (6,0) {$4$};
\node[SO] (E) at (8,0) {$5$};
\node[Sp] (F) at (8.5,0) {$4$};
\node[SU] (G) at (10.5,0) {$2$};
\draw[dashed] (-1,0)--(A);
\draw (A)--(B);
\draw[->] (B)-- node[above, pos=0.9]{$\spinor$} (C);
\draw[->] (C)--(D);
\draw[->] (D)--(E);
\draw[->] (F)--(G);
\end{tikzpicture}\ ,

\begin{tikzpicture}
\node[SO] (A) at (0,0) {$8$};
\node[Sp] (B) at (2,0) {$6$};
\node[SO] (C) at (4,0) {$7$};
\node[Sp] (D) at (6,0) {$4$};
\node[SO] (E) at (6.5,0) {$5$};
\node[Sp] (F) at (8.5,0) {$2$};
\node[SO] (G) at (10.5,0) {$3$};
\draw[dashed] (-1,0)--(A);
\draw (A)--(B);
\draw[->] (B)-- node[above, pos=0.9]{$\spinor$} (C);
\draw[->] (C)--(D);
\draw[->] (E)--(F);
\draw[->] (F)--(G);
\end{tikzpicture}\ ,

\begin{tikzpicture}
\node[SO] (A) at (0,0) {$8$};
\node[Sp] (B) at (2,0) {$6$};
\node[SO] (C) at (4,0) {$7$};
\node[Sp] (D) at (6,0) {$4$};
\node[SO] (E) at (8,0) {$4$};
\draw[dashed] (-1,0)--(A);
\draw (A)--(B);
\draw[->] (B)-- node[above, pos=0.9]{$\spinor$} (C);
\draw[->] (C)--(D);
\draw[->] (D)--(E);
\end{tikzpicture}\ ,

\begin{tikzpicture}
\node[SO] (A) at (0,0) {$8$};
\node[Sp] (B) at (2,0) {$6$};
\node[SO] (C) at (4,0) {$7$};
\node[Sp] (D) at (6,0) {$4$};
\node[SU] (E) at (8,0) {$2$};
\draw[dashed] (-1,0)--(A);
\draw (A)--(B);
\draw[->] (B)-- node[above, pos=0.9]{$\spinor$} (C);
\draw[->] (C)--(D);
\draw[->] (D)--(E);
\end{tikzpicture}\ ,

\begin{tikzpicture}
\node[SO] (A) at (0,0) {$8$};
\node[Sp] (B) at (2,0) {$6$};
\node[SO] (C) at (4,0) {$7$};
\node[Sp] (D) at (6,-1) {$2$};
\node[Sp] (P) at (6,1) {$2$};
\draw[dashed] (-1,0)--(A);
\draw (A)--(B);
\draw[->] (B)-- node[above, pos=0.9]{$\spinor$} (C);
\draw[->] (C)--(D);
\draw[->] (C)--(P);
\end{tikzpicture}\ .

\noindent In the last two branches, all the $\half\fund\otimes\fund$ of $\SO(7)\times\USp(4)$ or $\SO(7)\times\USp(2)$ can also be replaced by the corresponding $\half\spinor\otimes\fund$.

If the branch current $\leq1$, we can have \[
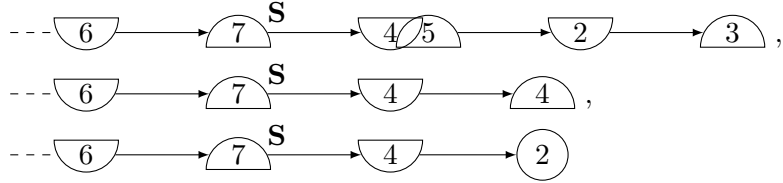
\begin{array}{l}
\begin{tikzpicture}
\node[Sp] (A) at (0,0) {$6$};
\node[SO] (B) at (2,0) {$7$};
\node[Sp] (C) at (4,0) {$4$};
\node[SO] (CC) at (4.5,0) {$5$};
\node[Sp] (D) at (6.5,0) {$2$};
\node[SO] (E) at (8.5,0) {$3$};
\draw[dashed] (-1,0)--(A);
\draw[->] (A)--(B);
\draw[->] (B)-- node[above, pos=0.1]{$\spinor$} (C);
\draw[->] (CC)--(D);
\draw[->] (D)--(E);
\end{tikzpicture}\ ,\\
\begin{tikzpicture}
\node[Sp] (A) at (0,0) {$6$};
\node[SO] (B) at (2,0) {$7$};
\node[Sp] (C) at (4,0) {$4$};
\node[SO] (D) at (6,0) {$4$};
\draw[dashed] (-1,0)--(A);
\draw[->] (A)--(B);
\draw[->] (B)-- node[above, pos=0.1]{$\spinor$} (C);
\draw[->] (C)--(D);
\end{tikzpicture}\ ,\\
\begin{tikzpicture}
\node[Sp] (A) at (0,0) {$6$};
\node[SO] (B) at (2,0) {$7$};
\node[Sp] (C) at (4,0) {$4$};
\node[SU] (D) at (6,0) {$2$};
\draw[dashed] (-1,0)--(A);
\draw[->] (A)--(B);
\draw[->] (B)-- node[above, pos=0.1]{$\spinor$} (C);
\draw[->] (C)--(D);
\end{tikzpicture}
\end{array}
\]
and \[
\begin{tikzpicture}
\node[SO] (A) at (0,0) {$9$};
\node[Sp] (B) at (2,0) {$6$};
\node (C) at (4,0) {$\mathsf{G}_2$};
\node[Sp] (D) at (6,0) {$2$};
\draw[dashed] (-1,0)--(A);
\draw[->] (A)--(B);
\draw (B)-- (C);
\draw (C)-- (D);
\end{tikzpicture}\ .
\] 

If the branch current $\leq2$, we can have:
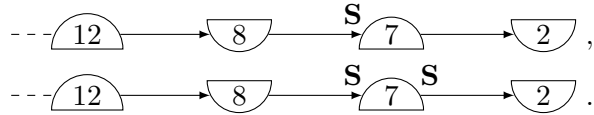
\[
\begin{array}{l}
\begin{tikzpicture}
\node[SO] (A) at (0,0) {$12$};
\node[Sp] (B) at (2,0) {$8$};
\node[SO] (C) at (4,0) {$7$};
\node[Sp] (D) at (6,0) {$2$};
\draw[dashed] (-1,0)--(A);
\draw[->] (A)--(B);
\draw[->] (B)-- node[above, pos=0.9]{$\spinor$} (C);
\draw[->] (C)--(D);
\end{tikzpicture}\ , \\ 
\begin{tikzpicture}
\node[SO] (A) at (0,0) {$12$};
\node[Sp] (B) at (2,0) {$8$};
\node[SO] (C) at (4,0) {$7$};
\node[Sp] (D) at (6,0) {$2$};
\draw[dashed] (-1,0)--(A);
\draw[->] (A)--(B);
\draw[->] (B)-- node[above, pos=0.9]{$\spinor$} (C);
\draw[->] (C)--node[above, pos=0.1]{$\spinor$}(D);
\end{tikzpicture}\ .
\end{array}
\]

If the branch current $\leq3$, we can have:\[
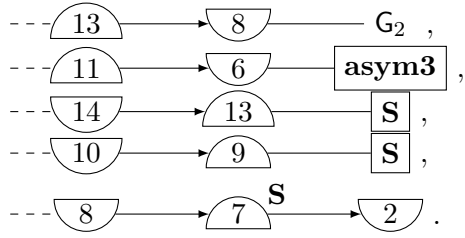
\begin{array}{l}
\begin{tikzpicture}
\node[SO] (A) at (0,0) {$13$};
\node[Sp] (B) at (2,0) {$8$};
\node (C) at (4,0) {$\mathsf{G}_2$};
\draw[dashed] (-1,0)--(A);
\draw[->] (A)--(B);
\draw (B)-- (C);
\end{tikzpicture}\ , \\
\begin{tikzpicture}
\node[SO] (A) at (0,0) {$11$};
\node[Sp] (B) at (2,0) {$6$};
\node[1gon] (C) at (4,0) {$\asymT$};
\draw[dashed] (-1,0)--(A);
\draw[->] (A)--(B);
\draw (B)--(C);
\end{tikzpicture}\ ,\\
\begin{tikzpicture}
\node[Sp] (A) at (0,0) {$14$};
\node[SO] (B) at (2,0) {$13$};
\node[1gon] (C) at (4,0) {$\spinor$};
\draw[dashed] (-1,0)--(A);
\draw[->] (A)--(B);
\draw (B)--(C);
\end{tikzpicture}\ , \\
\begin{tikzpicture}
\node[Sp] (A) at (0,0) {$10$};
\node[SO] (B) at (2,0) {$9$};
\node[1gon] (C) at (4,0) {$\spinor$};
\draw[dashed] (-1,0)--(A);
\draw[->] (A)--(B);
\draw (B)--(C);
\end{tikzpicture}\ ,\\
\begin{tikzpicture}
\node[Sp] (A) at (0,0) {$8$};
\node[SO] (B) at (2,0) {$7$};
\node[Sp] (C) at (4,0) {$2$};
\draw[dashed] (-1,0)--(A);
\draw[->] (A)--(B);
\draw[->] (B)-- node[above, pos=0.1]{$\spinor$} (C);
\end{tikzpicture}\ .
\end{array}
\]

If the branch current $\leq4$, we can have:\[
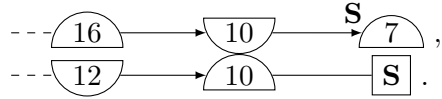
\begin{array}{l}
\begin{tikzpicture}
\node[SO] (A) at (0,0) {$16$};
\node[Sp] (B) at (2,0) {$10$};
\node[SO] (C) at (4,0) {$7$};
\draw[dashed] (-1,0)--(A);
\draw[->] (A)--(B);
\draw[->] (B)-- node[above, pos=0.9]{$\spinor$} (C);
\end{tikzpicture}\ ,\\
\begin{tikzpicture}
\node[Sp] (A) at (0,0) {$12$};
\node[SO] (B) at (2,0) {$10$};
\node[1gon] (C) at (4,0) {$\spinor$};
\draw[dashed] (-1,0)--(A);
\draw[->] (A)--(B);
\draw (B)--(C);
\end{tikzpicture}\ .
\end{array}
\]

If the branch current $\leq5$, we can have:\[
\begin{tikzpicture}
\node[Sp] (A) at (0,0) {$14$};
\node[SO] (B) at (2,0) {$11$};
\node[1gon] (C) at (4,0) {$\spinor$};
\draw[dashed] (-1,0)--(A);
\draw[->] (A)--(B);
\draw (B)--(C);
\end{tikzpicture}\ .
\]

If the branch current $\leq6$, we can have: \[
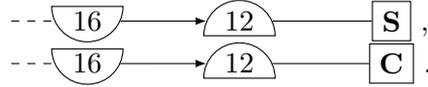
\begin{array}{l}
\begin{tikzpicture}
\node[Sp] (A) at (0,0) {$16$};
\node[SO] (B) at (2,0) {$12$};
\node[1gon] (C) at (4,0) {$\spinor$};
\draw[dashed] (-1,0)--(A);
\draw[->] (A)--(B);
\draw (B)--(C);
\end{tikzpicture}\ ,\\
\begin{tikzpicture}
\node[Sp] (A) at (0,0) {$16$};
\node[SO] (B) at (2,0) {$12$};
\node[1gon] (C) at (4,0) {$\conj$};
\draw[dashed] (-1,0)--(A);
\draw[->] (A)--(B);
\draw (B)--(C);
\end{tikzpicture}\ .
\end{array}
\]

\paragraph{Further branching:}
If the branch current $=0$, we can have: 
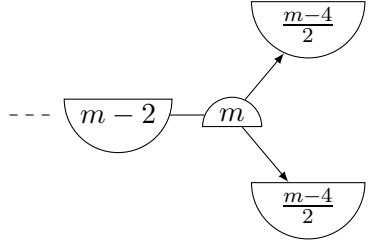
\[
\begin{tikzpicture}
\node[SO] (A) at  (-1,0)  {$m$};
\node[Sp] (B) at  (-2.5,0)  {$m-2$};
\node[Sp] (P) at  (0,1.2)  {$\frac{m-4}{2}$};
\node[Sp] (Q) at  (0,-1.2)  {$\frac{m-4}{2}$};
\draw (A)--(B);
\draw[<-] (P)--(A);
\draw[<-] (Q)--(A);
\draw[dashed] (-3.4,0)--(-4,0);
\end{tikzpicture}
\]for $m=4k, k\geq2$. For $m=12$, one or two $\USp(\frac{m-4}{2})$ nodes can be replaced by $\spinor$ of $\SO(12)$;
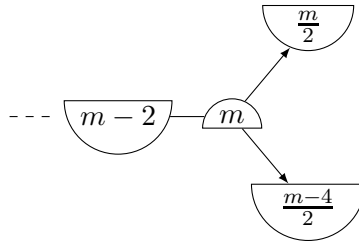
\[
\begin{tikzpicture}
\node[SO] (A) at  (-1,0)  {$m$};
\node[Sp] (B) at  (-2.5,0)  {$m-2$};
\node[Sp] (P) at  (0,1.2)  {$\frac{m}{2}$};
\node[Sp] (Q) at  (0,-1.2)  {$\frac{m-4}{2}$};
\draw (A)--(B);
\draw[<-] (P)--(A);
\draw[<-] (Q)--(A);
\draw[dashed] (-3.4,0)--(-4,0);
\end{tikzpicture}
\] for $m=4k, k\geq2$. For $m=12$, the $\USp(\frac{m-4}{2})$ node can be replaced by $\spinor$ of $\SO(12)$;
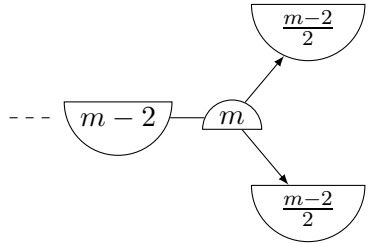
\[
\begin{tikzpicture}
\node[SO] (A) at  (-1,0)  {$m$};
\node[Sp] (B) at  (-2.5,0)  {$m-2$};
\node[Sp] (P) at  (0,1.2)  {$\frac{m-2}{2}$};
\node[Sp] (Q) at  (0,-1.2)  {$\frac{m-2}{2}$};
\draw (A)--(B);
\draw[<-] (P)--(A);
\draw[<-] (Q)--(A);
\draw[dashed] (-3.4,0)--(-4,0);
\end{tikzpicture}
\] for $m=4k+2, k\geq1$. For $m=10$, one or two $\USp(\frac{m-2}{2})$ nodes can be replaced by $\spinor$ of $\SO(10)$.

When $m=8$, there are four more possibilities: \[
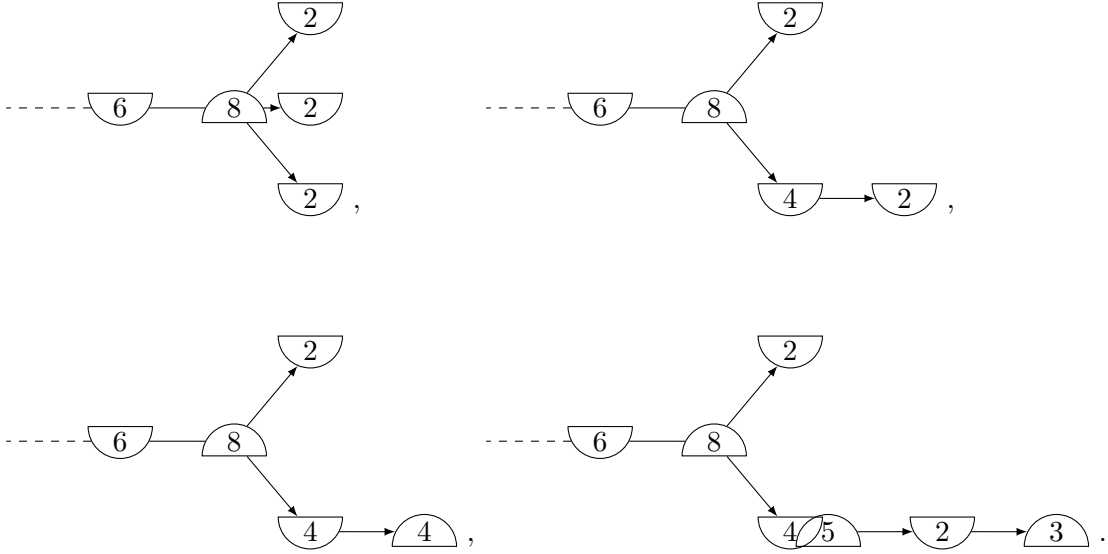
\begin{array}{ll}
\begin{tikzpicture}[baseline=(Q.base)]
\node[SO] (A) at  (-1,0)  {$8$};
\node[Sp] (B) at  (-2.5,0)  {$6$};
\node[Sp] (P) at  (0,1.2)  {$2$};
\node[Sp] (Q) at  (0,-1.2)  {$2$};
\node[Sp] (R) at  (0,0)  {$2$};
\draw (A)--(B);
\draw[<-] (P)--(A);
\draw[<-] (Q)--(A);
\draw[<-] (R)--(A);
\draw[dashed] (B)--(-4,0);
\end{tikzpicture}\ , &
\begin{tikzpicture}[baseline=(Q.base)]
\node[SO] (A) at  (-1,0)  {$8$};
\node[Sp] (B) at  (-2.5,0)  {$6$};
\node[Sp] (P) at  (0,1.2)  {$2$};
\node[Sp] (Q) at  (0,-1.2)  {$4$};
\node[Sp] (R) at  (1.5,-1.2)  {$2$};
\draw (A)--(B);
\draw[<-] (P)--(A);
\draw[<-] (R)--(Q);
\draw[<-] (Q)--(A);
\draw[dashed] (B)--(-4,0);
\end{tikzpicture}\ , \\[4em]
\begin{tikzpicture}[baseline=(Q.base)]
\node[SO] (A) at  (-1,0)  {$8$};
\node[Sp] (B) at  (-2.5,0)  {$6$};
\node[Sp] (P) at  (0,1.2)  {$2$};
\node[Sp] (Q) at  (0,-1.2)  {$4$};
\node[SO] (R) at  (1.5,-1.2)  {$4$};
\draw (A)--(B);
\draw[<-] (P)--(A);
\draw[<-] (R)--(Q);
\draw[<-] (Q)--(A);
\draw[dashed] (B)--(-4,0);
\end{tikzpicture}\ , &
\begin{tikzpicture}[baseline=(Q.base)]
\node[SO] (A) at  (-1,0)  {$8$};
\node[Sp] (B) at  (-2.5,0)  {$6$};
\node[Sp] (P) at  (0,1.2)  {$2$};
\node[Sp] (Q) at  (0,-1.2)  {$4$};
\node[SO] (QQ) at  (.5,-1.2)  {$5$};
\node[Sp] (R) at  (2,-1.2)  {$2$};
\node[SO] (S) at  (3.5,-1.2)  {$3$};
\draw (A)--(B);
\draw[<-] (P)--(A);
\draw[<-] (R)--(QQ);
\draw[<-] (S)--(R);
\draw[<-] (Q)--(A);
\draw[dashed] (B)--(-4,0);
\end{tikzpicture}\ .\end{array}
\]

If the branch current $\leq1$, we can have:
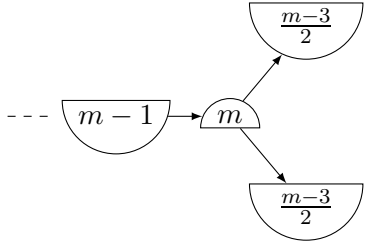
\[
\begin{tikzpicture}
\node[SO] (A) at  (-1,0)  {$m$};
\node[Sp] (B) at  (-2.5,0)  {$m-1$};
\node[Sp] (P) at  (0,1.2)  {$\frac{m-3}{2}$};
\node[Sp] (Q) at  (0,-1.2)  {$\frac{m-3}{2}$};
\draw[<-] (A)--(B);
\draw[<-] (P)--(A);
\draw[<-] (Q)--(A);
\draw[dashed] (-3.4,0)--(-4,0);
\end{tikzpicture}
\] for $m=4k+3, k\geq1$. For $m=11$, one or two $\USp(\frac{m-3}{2})$ nodes can be replaced by $\spinor$ of $\SO(11)$. For $m=7$, one or two $\half\fund\otimes\fund$ of $\SO(7)\times\USp(2)$ can be replaced by corresponding $\half\spinor\otimes\fund$;

If the branch current $\leq2$, we can have: 
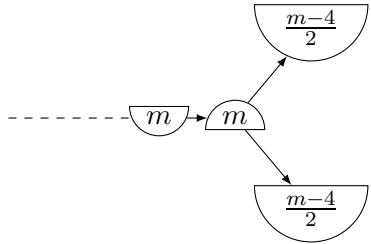
\[
\begin{tikzpicture}
\node[SO] (A) at  (-1,0)  {$m$};
\node[Sp] (B) at  (-2,0)  {$m$};
\node[Sp] (P) at  (0,1.2)  {$\frac{m-4}{2}$};
\node[Sp] (Q) at  (0,-1.2)  {$\frac{m-4}{2}$};
\draw[<-] (A)--(B);
\draw[<-] (P)--(A);
\draw[<-] (Q)--(A);
\draw[dashed] (-2.4,0)--(-4,0);
\end{tikzpicture}
\] for $m=4k, k\geq2$. For $m=12$, one or two $\USp(\frac{m-4}{2})$ nodes can be replaced by $\spinor$ of $\SO(12)$.

\paragraph{Branches involving identifications}
Identifications $\SO(6)\simeq \SU(4)$ and $\SO(5)\simeq \USp(4)$ introduce additional possibilities in the branches, which we enumerate here.

For a $\USp(6)$-$\SO(8)$ or $\SO(8)$-$\USp(6)$ branch with branch current zero, we have
\[
\begin{tikzpicture}
\node[Sp] (Z) at  (-1.5,0)  {6};
\node[SU] (A) at  (0,0)  {4};
\node[SO] (B) at  (.5,0)  {6};
\node[Sp] (C) at  (2,0)  {2};
\draw[->] (Z)--(A);
\draw[->] (B)--(C);
\draw[dashed] (Z)--(-2.5,0);
\end{tikzpicture}\ .
\]

For a $\USp(4)$-$\SO(6)$ or $\SO(6)$-$\USp(4)$ branch with branch current $=0$,
we have
\[
\begin{array}{l@{\qquad}l}
\begin{tikzpicture}
\node[SO] (A) at  (0,0)  {6};
\node[SU] (B) at  (.5,0)  {4};
\node[SU] (C) at  (2,-.8)  {2};
\node[SU] (D) at  (2,.8)  {2};
\draw[->] (B)--(C);
\draw[->] (B)--(D);
\draw[dashed] (A)--(-1,0);
\end{tikzpicture}\ ,&
\begin{tikzpicture}
\node[SO] (A) at  (0,0)  {6};
\node[SU] (B) at  (.5,0)  {4};
\node[Sp] (C) at  (2,0)  {4};
\node[SU] (D) at  (3.5,0)  {2};
\draw (B)--(C);
\draw[->] (C)--(D);
\draw[dashed] (A)--(-1,0);
\end{tikzpicture}\ ,\\
\begin{tikzpicture}
\node[SO] (A) at  (0,0)  {6};
\node[SU] (B) at  (.5,0)  {4};
\node[Sp] (C) at  (2,0)  {4};
\node[SO] (D) at  (3.5,0)  {4};
\draw (B)--(C);
\draw[->] (C)--(D);
\draw[dashed] (A)--(-1,0);
\end{tikzpicture}\ ,&
\begin{tikzpicture}
\node[SO] (A) at  (0,0)  {6};
\node[SU] (B) at  (.5,0)  {4};
\node[Sp] (C) at  (2,0)  {4};
\node[SO] (CC) at  (2.5,0)  {5};
\node[Sp] (D) at  (4,0)  {2};
\node[SO] (E) at  (5.5,0)  {3};
\draw (B)--(C);
\draw[->] (CC)--(D);
\draw[->] (D)--(E);
\draw[dashed] (A)--(-1,0);
\end{tikzpicture}\ .
\end{array}
\]

For a $\USp$-$\SO$ or $\SO$-$\USp$ branch with branch current $\leq1$,
\[
\begin{tikzpicture}
\node[SO] (A) at  (0,0)  {5};
\node[Sp] (B) at  (.5,0)  {4};
\node[SO] (C) at  (2,0)  {4};
\draw[->] (B)--(C);
\draw[dashed] (A)--(-1,0);
\end{tikzpicture}\ , 
\qquad
\begin{tikzpicture}
\node[SO] (A) at  (0,0)  {5};
\node[Sp] (B) at  (.5,0)  {4};
\node[SU] (C) at  (2,0)  {2};
\draw[->] (B)--(C);
\draw[dashed] (A)--(-1,0);
\end{tikzpicture}\ .
\]

For a $\USp$-$\SO$ or $\SO$-$\USp$ branch with branch current $\leq2$, \[
\begin{tikzpicture}
\node[SO] (A) at  (0,0)  {6};
\node[SU] (B) at  (.5,0)  {4};
\node[SU] (C) at  (2,0)  {2};
\draw[->] (B)--(C);
\draw[dashed] (A)--(-1,0);
\end{tikzpicture}\ , \qquad
\begin{tikzpicture}
\node[Sp] (A) at  (0,0)  {4};
\node[SO] (B) at  (.5,0)  {5};
\node[Sp] (C) at  (2,0)  {2};
\node[SO] (D) at  (3.5,0)  {3};
\draw[->] (B)--(C);
\draw[->] (C)--(D);
\draw[dashed] (A)--(-1,0);
\end{tikzpicture}
\]

\subsubsection{$\USp$-$\SU$ branches (usually large)}
Notice that, because the current $I$ in the first edge of the branch must be non-negative and the beta function condition for $\USp$ reduces to $I+2\geq0$, we can obtain every $\USp-\SU$ branch by replacing the zero-th vertex $\SU(m_0)$ by $\USp(m_0)$ in a $\SU-\SU$ branch provided that $m_0$ is even. These branches are usually large. The branch current is bigger than or equal to $-(m_0-2)$. 

\noindent These are small only when \[
\begin{tikzpicture}
\node[Sp] (0) at  (-2,0)  {$2m$};
\node[SU] (A) at  (0,0)  {$m$};
\draw[->] (0)--(A);
\end{tikzpicture}
\] 
with branch current 2,
and \[
\begin{tikzpicture}
\node[Sp] (0) at  (-2,0)  {$2m$};
\node[SU] (A) at  (0,0)  {$m+1$};
\draw[->] (0)--(A);
\end{tikzpicture}
\]
with branch current 0.  The latter can be extended in two cases:
\[
\begin{tikzpicture}
\node[Sp] (B) at (2,0) {$6$};
\node[SU] (C) at (4,0) {$4$};
\node[SU] (D) at (6,0) {$2$};
\draw[->] (B)--(C);
\draw[->] (C)--(D);
\end{tikzpicture}\ , \qquad
\begin{tikzpicture}
\node[Sp] (B) at (2,0) {$4$};
\node[SU] (C) at (4,0) {$3$};
\node[SU] (D) at (6,0) {$2$};
\draw[->] (B)--(C);
\draw[->] (C)--(D);
\end{tikzpicture}\ .
\] 

\subsubsection{ $\USp$-$\USp$  branches (usually large)}

These are of two types.
Both are a degenerate case of a $\USp$-$\SU$ branch.
 The one is \[
\begin{tikzpicture}
\node[Sp] (B) at  (2,0)  {$m$};
\node[Sp] (C) at  (3.5,0)  {$m$};
\draw (B)--(C);
\end{tikzpicture}
\] with the branch current $-(m-2)$. This is always large. The second is \[
\begin{tikzpicture}
\node[Sp] (B) at  (2,0)  {$m$};
\node[Sp] (C) at  (3.5,0)  {$m-2$};
\draw[->] (B)--(C);
\end{tikzpicture}
\] with the branch current  $-(m-6)$. This is usually large.

 The first type of branch can be extended only in one case: 
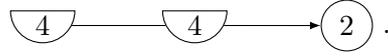
\[
\begin{tikzpicture}
\node[Sp] (B) at  (2,0)  {$4$};
\node[Sp] (C) at  (4,0)  {$4$};
\node[SU] (D) at  (6,0)  {$2$};
\draw (B)--(C);
\draw[->] (C)--(D);
\end{tikzpicture}\ .
\]
The second type of branches are small only when $m=4$ or $m=6$.  For $m=4$ it reduces to $\USp-\SU$ branch. For $m=6$ the beta function contribution is exactly half the maximum allowed.

We also have the decorations of the form \[
\begin{tikzpicture}
\node[Sp] (B) at  (2,0)  {$m$};
\node[1gon] (C) at  (3.5,0)  {$\asym$};
\draw (B)--(C);
\end{tikzpicture} \]
which behave numerically the same as the second type of branches above.
These are usually large with branch current $-(m-6)$. They are small only when $m=4$ or $m=6$. For $m=4$, it is the same as $\vect$ of $\SO(5)$ and treated as such.

\subsubsection{$\SO$-$\SU$  branches (always large)}
Notice that if we take a $\SU-\SU$ branch with branch current $\geq2$, then we can replace $\SU(m_0)$ by $\SO(m_0)$ to obtain a $\SO-\SU$ branch. Every $\SO-\SU$ branch may be obtained this way. The apparent current satisfies $I_0\ge 2$. The branch current $I\geq -(m_0-2)$ which means these branches are always large.

\subsubsection{$\mathsf{G}_2$  branches}
They are all obtained by taking $\SO(7)-\USp(2m)$ branches and replacing the zero-th $\SO(7)$ node with $\mathsf{G}_2$.

\subsection{Main classification}\label{combinations}

Having classified the possible types of trunks and branches, we turn to the problem of classifying various ways of combining trunks and branches together.  As we will see, except for a finite number of gauge theories, the associated graph is either finite or affine Dynkin diagram. 
As was discussed in Sec.~\ref{overall}, the choices are \begin{itemize}
\item There is a single loop.
\item There is a trunk with more than one node. In this case we can only attach small branches on both ends. 
\item There is a trunk with only one node. 
\end{itemize}
We study them in turn. Before proceeding, we note that $\spinor$, $\conj$ and $\vect$ of $\SO(8)$ are not separately listed in the classification. 

\subsubsection{Theories with a single loop}\label{loop}
The possibilities are 
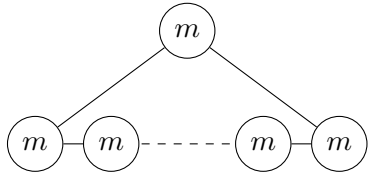
\[
\begin{tikzpicture}
\node[SU] (A) at  (4,1.5)  {$m$};
\node[SU] (B) at  (2,0)  {$m$};
\node[SU] (C) at  (3,0)  {$m$};
\node[SU] (D) at  (5,0)  {$m$};
\node[SU] (E) at  (6,0)  {$m$};
\draw (A)--(B)--(C);
\draw (A)--(E);
\draw[dashed] (3.4,0)--(4.6,0);
\draw (D)--(E);
\end{tikzpicture}
\] or 
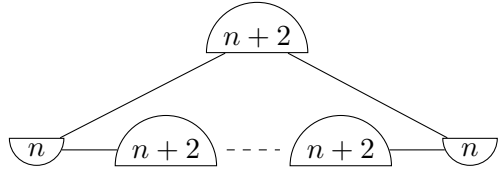
\[
\begin{tikzpicture}
\node[Sp] (A) at  (0.8,0)  {$n$};
\node[SO] (B) at  (2.5,0)  {$n+2$};
\node[SO] (C) at  (4.8,0)  {$n+2$};
\node[Sp] (D) at  (6.5,0)  {$n$};
\node[SO] (P) at  (3.7,1.5)  {$n+2$};
\draw (A)--(B);
\draw (C)--(D);
\draw	(A)--(P)--(D);
\draw[dashed] (3.3,0)--(4,0);
\end{tikzpicture}
\] 
or a mixture of these when they involve $\SU(4)$ and $\SO(6)$. One example is 
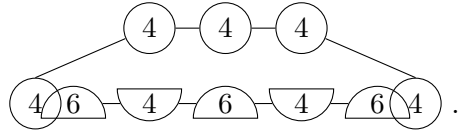
\[
	\begin{tikzpicture}
\node[SU] (A) at  (5.5,0)  {$4$};
\node[SU] (B) at  (4,1)  {$4$};
\node[SU] (C) at  (3,1)  {$4$};
\node[SU] (D) at  (2,1)  {$4$};
\node[SU] (E) at  (.5,0)  {$4$};
\node[SO] (P) at  (1,0)  {$6$};
\node[Sp] (Q) at  (2,0)  {$4$};
\node[SO] (R) at  (3,0)  {$6$};
\node[Sp] (S) at  (4,0)  {$4$};
\node[SO] (T) at  (5,0)  {$6$};
\draw (A.north)--(B)--(C)--(D)--(E.north);
\draw (P)--(Q)--(R)--(S)--(T);
\end{tikzpicture}\ .
\]

We pause here to note that in the classification, we basically neglected the distinction between a $\fund\otimes\fund$ and a $\fund\otimes\overline{\fund}$ of $\SU(n)\times \SU(m)$. When the $\SU$ nodes form a tree, by an appropriate application of the outer automorphism at each $\SU$ node, we can make every such bi-fundamentals to be $\fund\otimes\overline{\fund}$.
However, when the associated graph is a loop consisting purely of $\SU(m)$, this process can fail at one node, and therefore there are \emph{two} distinct theories whose associated graph is a loop of $\SU(m)$. 

A similar comment must be made here for the $\SO(8)$-$\USp(6)$ ``bifundamentals'' here, which can stand for either $\half\vect\otimes \vect$, $\half\spinor\otimes \vect$, or $\half\conj\otimes \vect$. At each $\SO(8)$ node, the only information invariant under outer automorphism is whether the two ``bifundamentals'' on both sides of $\SO(8)$ are the same or not.  We do not repeat similar statements below.

\subsubsection{Theories whose trunk has more than one node}\label{tree-with-a-long-trunk}

\paragraph{$\SU$ trunk}
Given a trunk of the form 
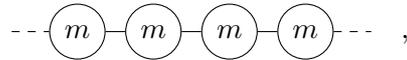
\[
\begin{tikzpicture}
\node (0) at (0,0) {};
\node[SU] (A) at (1,0) {$m$};
\node[SU] (B) at (2,0) {$m$};
\node[SU] (C) at (3,0) {$m$};
\node[SU] (D) at (4,0) {$m$};
\node (E) at (5,0) {};
\draw[dashed] (0)--(A);
\draw[dashed] (D)--(E);
\draw (A)--(B)--(C)--(D);
\end{tikzpicture}\ ,
\]
 we can attach any $\SU$ branches at both the ends of the trunk. This gives rise to A, D type of finite Dynkin graphs and D type of affine Dynkin graphs. The decorations occur only in the branches.

\paragraph{$\SO-\USp$ trunk}
Given a trunk of the form
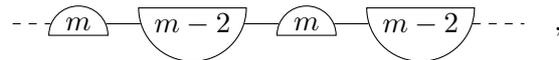
\[
\begin{tikzpicture}
\node (0) at (0,0) {};
\node[SO] (A) at (1,0) {$m$};
\node[Sp] (B) at (2.5,0) {$m-2$};
\node[SO] (C) at (4,0) {$m$};
\node[Sp] (D) at (5.5,0) {$m-2$};
\node (E) at (7,0) {};
\draw[dashed] (0)--(A);
\draw[dashed] (D)--(E);
\draw (A)--(B)--(C)--(D);
\end{tikzpicture}\ ,
\]
we can attach any small $\SO$ or $\USp$  branches at both the ends of the trunk. This gives rise to A, D type of finite Dynkin graphs and D type of affine Dynkin graphs with a few exceptional graphs. The basic constituents of these exceptional graphs are the following four branches: \[
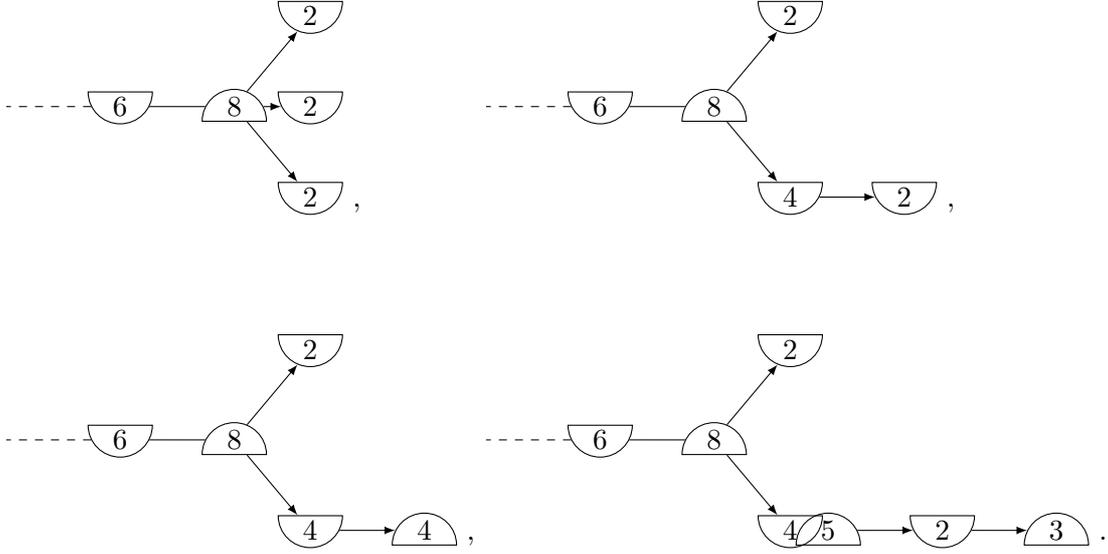
\begin{array}{ll}
\begin{tikzpicture}[baseline=(Q.base)]
\node[SO] (A) at  (-1,0)  {$8$};
\node[Sp] (B) at  (-2.5,0)  {$6$};
\node[Sp] (P) at  (0,1.2)  {$2$};
\node[Sp] (Q) at  (0,-1.2)  {$2$};
\node[Sp] (R) at  (0,0)  {$2$};
\draw (A)--(B);
\draw[<-] (P)--(A);
\draw[<-] (Q)--(A);
\draw[<-] (R)--(A);
\draw[dashed] (B)--(-4,0);
\end{tikzpicture}\ , &
\begin{tikzpicture}[baseline=(Q.base)]
\node[SO] (A) at  (-1,0)  {$8$};
\node[Sp] (B) at  (-2.5,0)  {$6$};
\node[Sp] (P) at  (0,1.2)  {$2$};
\node[Sp] (Q) at  (0,-1.2)  {$4$};
\node[Sp] (R) at  (1.5,-1.2)  {$2$};
\draw (A)--(B);
\draw[<-] (P)--(A);
\draw[<-] (R)--(Q);
\draw[<-] (Q)--(A);
\draw[dashed] (B)--(-4,0);
\end{tikzpicture}\ , \\[4em]
\begin{tikzpicture}[baseline=(Q.base)]
\node[SO] (A) at  (-1,0)  {$8$};
\node[Sp] (B) at  (-2.5,0)  {$6$};
\node[Sp] (P) at  (0,1.2)  {$2$};
\node[Sp] (Q) at  (0,-1.2)  {$4$};
\node[SO] (R) at  (1.5,-1.2)  {$4$};
\draw (A)--(B);
\draw[<-] (P)--(A);
\draw[<-] (R)--(Q);
\draw[<-] (Q)--(A);
\draw[dashed] (B)--(-4,0);
\end{tikzpicture}\ , &
\begin{tikzpicture}[baseline=(Q.base)]
\node[SO] (A) at  (-1,0)  {$8$};
\node[Sp] (B) at  (-2.5,0)  {$6$};
\node[Sp] (P) at  (0,1.2)  {$2$};
\node[Sp] (Q) at  (0,-1.2)  {$4$};
\node[SO] (QQ) at  (.5,-1.2)  {$5$};
\node[Sp] (R) at  (2,-1.2)  {$2$};
\node[SO] (S) at  (3.5,-1.2)  {$3$};
\draw (A)--(B);
\draw[<-] (P)--(A);
\draw[<-] (R)--(QQ);
\draw[<-] (S)--(R);
\draw[<-] (Q)--(A);
\draw[dashed] (B)--(-4,0);
\end{tikzpicture}\ .\end{array}
\]
These branches are attached to the alternating $\SO(8)-\USp(6)$ trunk. We can attach any possible branch at the one remaining end of this trunk to obtain an exceptional graph. Notice that most of these exceptional graphs are of non-Dynkin type. 

\paragraph{Mixed $\SU(4)$-$\SO(6)$-$\USp(4)$ trunk}
Finally, we can have trunks mixing $\SU(4)$ chains and $\SO(6)$-$\USp(4)$ chains. One example is
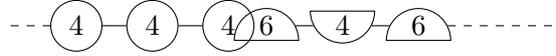
\[
	\begin{tikzpicture}
\node (G) at  (-1,0)  {};
\node[SU] (F) at  (0,0)  {$4$};
\node[SU] (A) at  (1,0)  {$4$};
\node[SU] (B) at  (2,0)  {$4$};
\node[SO] (C) at  (2.5,0)  {$6$};
\node[Sp] (D) at  (3.5,0)  {$4$};
\node[SO] (E) at  (4.5,0)  {$6$};
\node (Q) at  (6.5,0)  {};
\draw[dashed] (G)--(F);
\draw (F)--(A)--(B);
\draw (C)--(D)--(E);
\draw[dashed] (E)--(Q);
\end{tikzpicture} \]
where the two overlapping nodes  are again to be identified.
In this case, depending on the ends of the trunk being $\SU(4)$, $\SO(6)$ or $\USp(4)$,
we add corresponding small branches.

\subsubsection{$\SU$ trunk with exactly one node}\label{su-trunk}

\paragraph{5-valent or higher:} There are no such graphs.

\paragraph{4-valent:} From the local structure at a node, we see  that the center is $\SU(2m)$ and the branch current is always $m$. 
So, it is of affine D type. We can have decorations when $m=2,3$. When $m=2$, $\SU(2)$ nodes can be swapped with $\asym$, or when $m=3$, $\SU(3)$ nodes can be swapped with $\asymT$.

\paragraph{3-valent:} Call this central node $\SU(m)$. Call the branch currents in three directions as $I_1$, $I_2$ and $I_3$. The branch $i$ could have length of $k$ nodes (including the zero-th node $\SU(m)$) only if $0\leq I_i\leq \frac{m}{k}$. Also, we must have $I_1+I_2+I_3\geq m$. We can easily construct all the graphs having central node as $\SU(m)$ by taking three branches whose zero-th node is $\SU(m)$ and whose branch currents satisfy the above two inequalities. We now discuss some important properties of these graphs by dividing them into the following \emph{categories}:
\begin{itemize}
\item If the length of two branches is 2, then the length of the third branch is not restricted by any finite number for even $m$. When $m$ is even, and the current in the third branch is zero, we obtain the finite and affine D type Dynkin graphs which all have been counted earlier when we discussed the case of $\SU$ trunk having more than one node. If the current in the third branch is not zero, all the graphs are of finite D type. For odd $m$, the length of third branch cannot be more than $m-1$. All these graphs are of finite D type. There can be various decorations in the branches.
\item If the length of one branch is 2 and the length of another one is 3, then, from the above mentioned inequalities, we find that the length of third branch cannot be more than 6. When the length of the third branch is 6, the graph is an affine E$_8$ graph with a very restricted set of allowed labellings:
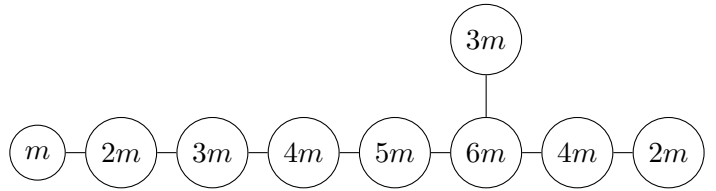
\[
\begin{tikzpicture}[baseline=(X.base)]
\node[SU] (X) at  (-1.5,0)  {$m$};
\node[SU] (Y) at  (-0.4,0)  {$2m$};
\node[SU] (Z) at  (0.8,0)  {$3m$};
\node[SU] (A) at  (4.4,1.5)  {$3m$};
\node[SU] (B) at  (2,0)  {$4m$};
\node[SU] (C) at  (3.2,0)  {$5m$};
\node[SU] (D) at  (4.4,0)  {$6m$};
\node[SU] (E) at  (5.6,0)  {$4m$};
\node[SU] (F) at  (6.8,0)  {$2m$};
\draw (A)--(D)--(E)--(F);
\draw (X)--(Y)--(Z)--(B)--(C)--(D);
\end{tikzpicture}\ .
\]
The rest of the graphs in this category are of finite E type. We can have various decorations in the branches.
\item If the length of two branches is 2 and another one is 4, then the length of third branch can at most be 4. All the graphs in which the length of third branch is less than 4 have already been counted above. When the length of the third branch is 4, the graph is an affine E$_7$ with a very restricted set of allowed labellings:
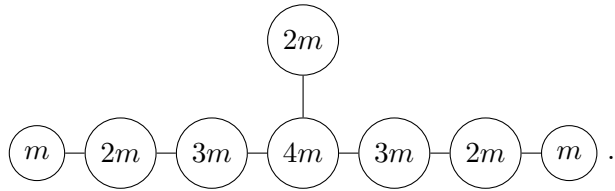
\[
\begin{tikzpicture}[baseline=(X.base)]
\node[SU] (X) at  (-1.5,0)  {$m$};
\node[SU] (Y) at  (-0.4,0)  {$2m$};
\node[SU] (Z) at  (0.8,0)  {$3m$};
\node[SU] (A) at  (2,1.5)  {$2m$};
\node[SU] (B) at  (2,0)  {$4m$};
\node[SU] (C) at  (3.2,0)  {$3m$};
\node[SU] (D) at  (4.4,0)  {$2m$};
\node[SU] (E) at  (5.5,0)  {$m$};
\draw (A)--(B)--(C)--(D)--(E);
\draw (X)--(Y)--(Z)--(B);
\end{tikzpicture}\ .
\]
\item If the length of two branches is 3, then the length of third branch is fixed to be 3. The graph is an affine E$_6$ with a very restricted set of allowed labellings:
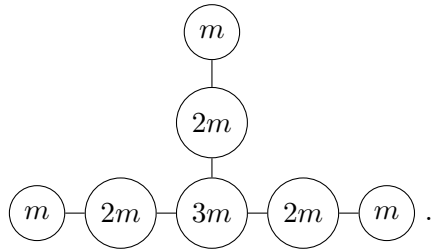
\[
\begin{tikzpicture}[baseline=(X.base)]
\node[SU] (X) at  (-1.5,0)  {$m$};
\node[SU] (Y) at  (-0.4,0)  {$2m$};
\node[SU] (Z) at  (0.8,0)  {$3m$};
\node[SU] (A) at  (0.8,1.2)  {$2m$};
\node[SU] (B) at  (0.8,2.4)  {$m$};
\node[SU] (D) at  (2,0)  {$2m$};
\node[SU] (E) at  (3.1,0)  {$m$};
\draw (B)--(A)--(Z)--(D)--(E);
\draw (X)--(Y)--(Z);
\end{tikzpicture}\ .
\]
\end{itemize}
\paragraph{2-valent:} To remove the graphs which have already been counted before, we may impose that the branch currents in both the directions are strictly positive. Then, all such graphs can be viewed as a degenerate case of the $\SU-\SU$ trunk with more than one nodes, and can be obtained by shrinking the trunk to exactly one node. All these graphs are of finite A type.

We also have cases involving large branches of length-1, 
listed in Sec.~\ref{what?}.
With $\sym$ of $\SU$, the possibilities are \[
\begin{tikzpicture}
\node[1gon] (A) at (0,0) {$\sym$};
\node[SU] (B) at (1.5,0) {$n$};
\node[SU] (C) at (3,0) {$m$};
\draw (A)--(B);
\draw[->] (B)--(C);
\draw[dashed] (C)--(4.5,0);
\end{tikzpicture} 
\] where $m\leq n-2$. In other words, we can add any $\SU$ branch with $I_0\ge 2$. 

With $\asymT$ of $\SU(7)$, the only possibility is 
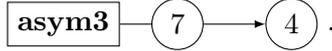
\[
\begin{tikzpicture}
\node[1gon] (A) at (0,0) {$\asymT$};
\node[SU] (B) at (1.5,0) {$7$};
\node[SU] (C) at (3,0) {$4$};
\draw (A)--(B);
\draw[->] (B)--(C);
\end{tikzpicture}\ .
\] A priori, we can add any branch with $I_0\ge 3$, but this turns out to be the only possible case.

\paragraph{1-valent:} These are precisely the branches with branch current $I>0$.

\paragraph{Comment:} Notice that all the graphs discussed in the above paragraphs turned out to be finite or affine Dynkin graphs. In particular, we find that the graphs corresponding to all $\SU-\SU$ bifundamentals are only Dynkin graphs, a fact which is already known in the literature and goes under the name \emph{ADE classification}.

\subsubsection{$\SO$ trunk with exactly one node}\label{so-trunk}

\paragraph{7-valent or higher:} There are no such graphs.

\paragraph{6-valent:} The only graph is 
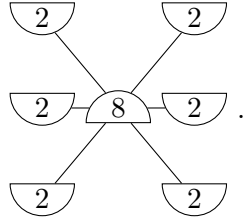
\[
\begin{tikzpicture}
\node[SO] (A) at  (1,0)  {8};
\node[Sp] (B) at  (0,1.2)  {2};
\node[Sp] (C) at  (0,0)  {2};
\node[Sp] (D) at  (2,1.2)  {2};
\node[Sp] (E) at  (2,0)  {2};
\node[Sp] (F) at  (2,-1.2)  {2};
\node[Sp] (G) at  (0,-1.2)  {2};
\draw (C)--(A)--(E);
\draw (B)--(A);
\draw (D)--(A);
\draw (F)--(A);
\draw (G)--(A);
\end{tikzpicture}\ .
\]
It is clearly a non-Dynkin type graph. This graph can also be viewed as a degenerate version (the trunk contracted to exactly one node) of a $\SO(8)-\USp(6)$ trunk with more than one node and closed on both ends by the first type of exceptional branch that we described above when we discussed the $\SO-\USp$ trunks with more than node.

\paragraph{5-valent:} The list is:
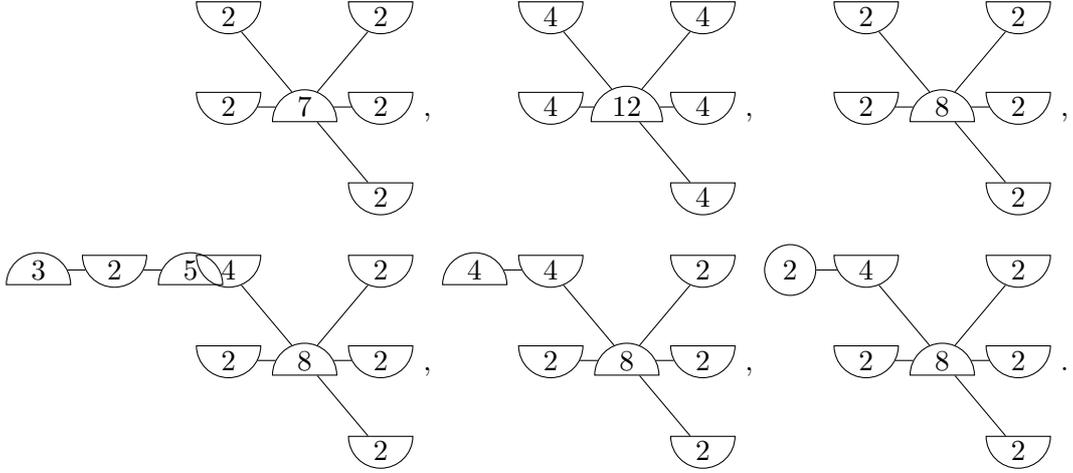
\[
\begin{array}{rrr}
\begin{tikzpicture}
\node[SO] (A) at  (1,0)  {7};
\node[Sp] (B) at  (0,1.2)  {2};
\node[Sp] (C) at  (0,0)  {2};
\node[Sp] (D) at  (2,1.2)  {2};
\node[Sp] (E) at  (2,0)  {2};
\node[Sp] (F) at  (2,-1.2)  {2};
\draw (C)--(A)--(E);
\draw (B)--(A);
\draw (D)--(A);
\draw (F)--(A);
\end{tikzpicture}\ , &
\begin{tikzpicture}
\node[SO] (A) at  (1,0)  {12};
\node[Sp] (B) at  (0,1.2)  {4};
\node[Sp] (C) at  (0,0)  {4};
\node[Sp] (D) at  (2,1.2)  {4};
\node[Sp] (E) at  (2,0)  {4};
\node[Sp] (F) at  (2,-1.2)  {4};
\draw (C)--(A)--(E);
\draw (B)--(A);
\draw (D)--(A);
\draw (F)--(A);
\end{tikzpicture}\ , &
\begin{tikzpicture}
\node[SO] (A) at  (1,0)  {8};
\node[Sp] (B) at  (0,1.2)  {2};
\node[Sp] (C) at  (0,0)  {2};
\node[Sp] (D) at  (2,1.2)  {2};
\node[Sp] (E) at  (2,0)  {2};
\node[Sp] (F) at  (2,-1.2)  {2};
\draw (C)--(A)--(E);
\draw (B)--(A);
\draw (D)--(A);
\draw (F)--(A);
\end{tikzpicture}\ , \\[4em]
\begin{tikzpicture}
\node[SO] (A) at  (1,0)  {8};
\node[Sp] (B) at  (0,1.2)  {4};
\node[SO] (BB) at  (-.5,1.2)  {5};
\node[Sp] (P) at  (-1.5,1.2)  {2};
\node[SO] (Q) at  (-2.5,1.2)  {3};
\node[Sp] (C) at  (0,0)  {2};
\node[Sp] (D) at  (2,1.2)  {2};
\node[Sp] (E) at  (2,0)  {2};
\node[Sp] (F) at  (2,-1.2)  {2};
\draw (C)--(A)--(E);
\draw (B)--(A);
\draw (Q)--(P)--(BB);
\draw (D)--(A);
\draw (F)--(A);
\end{tikzpicture}\ , &
\begin{tikzpicture}
\node[SO] (A) at  (1,0)  {8};
\node[Sp] (B) at  (0,1.2)  {4};
\node[SO] (P) at  (-1,1.2)  {4};
\node[Sp] (C) at  (0,0)  {2};
\node[Sp] (D) at  (2,1.2)  {2};
\node[Sp] (E) at  (2,0)  {2};
\node[Sp] (F) at  (2,-1.2)  {2};
\draw (C)--(A)--(E);
\draw (P)--(B)--(A);
\draw (D)--(A);
\draw (F)--(A);
\end{tikzpicture}\ ,&
\begin{tikzpicture}
\node[SO] (A) at  (1,0)  {8};
\node[Sp] (B) at  (0,1.2)  {4};
\node[SU] (P) at  (-1,1.2)  {2};
\node[Sp] (C) at  (0,0)  {2};
\node[Sp] (D) at  (2,1.2)  {2};
\node[Sp] (E) at  (2,0)  {2};
\node[Sp] (F) at  (2,-1.2)  {2};
\draw (C)--(A)--(E);
\draw (P)--(B)--(A);
\draw (D)--(A);
\draw (F)--(A);
\end{tikzpicture}\ .
\end{array}
\]
Some of them can be viewed as degenerate cases of graphs already counted. All these graphs are also of non-Dynkin type.

\paragraph{4-valent:} First, notice that any subgraph of an allowed graph is also an allowed graph. Hence, to obtain all graphs we can start with smaller graphs and try to extend them as far as we can. We adopt this strategy to obtain all allowed graphs having a 4-valent $\SO$ vertex. At the first level, when all branches are of length 2, we find that the only possibilities are:
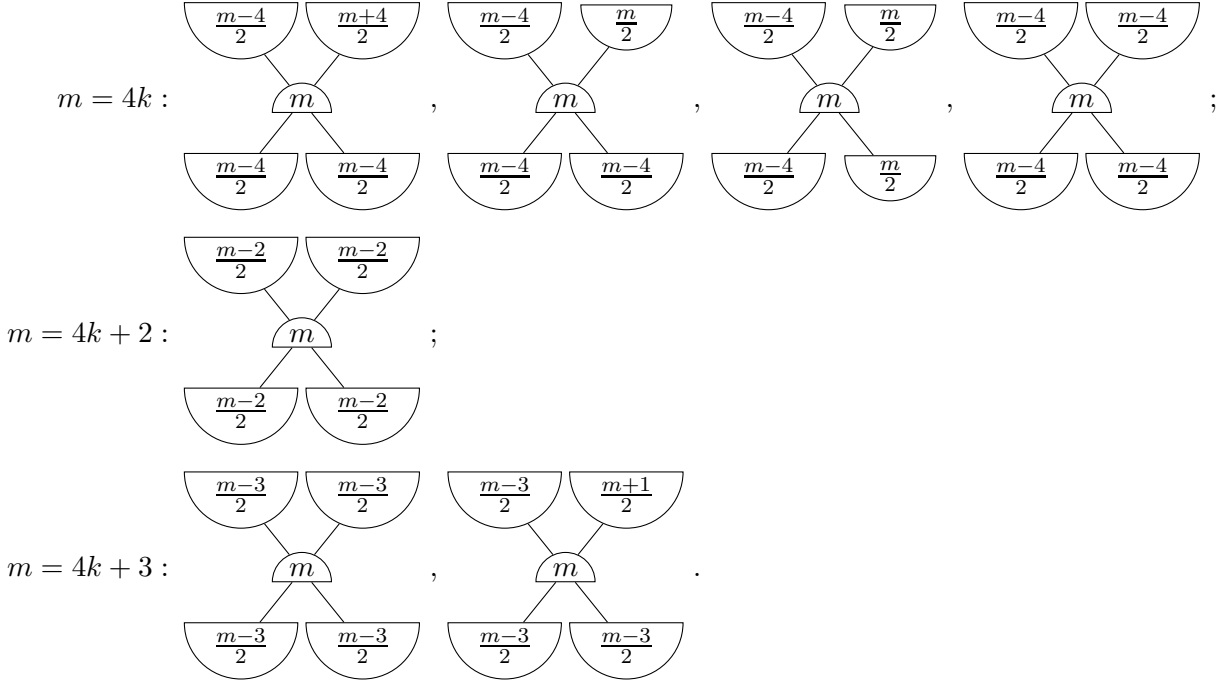
\begin{itemize}
\item[$m=4k:$]
\begin{tikzpicture}[xscale=.8]
\node[SO] (A) at  (1,0)  {$m$};
\node[Sp] (B) at  (0,1.0)  {$\frac{m-4}{2}$};
\node[Sp] (D) at  (2,1.0)  {$\frac{m+4}{2}$};
\node[Sp] (F) at  (2,-1.0)  {$\frac{m-4}{2}$};
\node[Sp] (G) at  (0,-1.0)  {$\frac{m-4}{2}$};
\draw (B)--(A);
\draw (D)--(A);
\draw (F)--(A);
\draw (G)--(A);
\end{tikzpicture}\ ,
\begin{tikzpicture}[xscale=.8]
\node[SO] (A) at  (1,0)  {$m$};
\node[Sp] (B) at  (0,1.0)  {$\frac{m-4}{2}$};
\node[Sp] (D) at  (2,1.0)  {$\frac{m}{2}$};
\node[Sp] (F) at  (2,-1.0)  {$\frac{m-4}{2}$};
\node[Sp] (G) at  (0,-1.0)  {$\frac{m-4}{2}$};
\draw (B)--(A);
\draw (D)--(A);
\draw (F)--(A);
\draw (G)--(A);
\end{tikzpicture}\ ,
\begin{tikzpicture}[xscale=.8]
\node[SO] (A) at  (1,0)  {$m$};
\node[Sp] (B) at  (0,1.0)  {$\frac{m-4}{2}$};
\node[Sp] (D) at  (2,1.0)  {$\frac{m}{2}$};
\node[Sp] (F) at  (2,-1.0)  {$\frac{m}{2}$};
\node[Sp] (G) at  (0,-1.0)  {$\frac{m-4}{2}$};
\draw (B)--(A);
\draw (D)--(A);
\draw (F)--(A);
\draw (G)--(A);
\end{tikzpicture}\ ,
\begin{tikzpicture}[xscale=.8]
\node[SO] (A) at  (1,0)  {$m$};
\node[Sp] (B) at  (0,1.0)  {$\frac{m-4}{2}$};
\node[Sp] (D) at  (2,1.0)  {$\frac{m-4}{2}$};
\node[Sp] (F) at  (2,-1.0)  {$\frac{m-4}{2}$};
\node[Sp] (G) at  (0,-1.0)  {$\frac{m-4}{2}$};
\draw (B)--(A);
\draw (D)--(A);
\draw (F)--(A);
\draw (G)--(A);
\end{tikzpicture}\ ; 
\item[$m=4k+2:$] 
\begin{tikzpicture}[xscale=.8]
\node[SO] (A) at  (1,0)  {$m$};
\node[Sp] (B) at  (0,1.0)  {$\frac{m-2}{2}$};
\node[Sp] (D) at  (2,1.0)  {$\frac{m-2}{2}$};
\node[Sp] (F) at  (2,-1.0)  {$\frac{m-2}{2}$};
\node[Sp] (G) at  (0,-1.0)  {$\frac{m-2}{2}$};
\draw (B)--(A);
\draw (D)--(A);
\draw (F)--(A);
\draw (G)--(A);
\end{tikzpicture}\ ; 
\item[$m=4k+3:$]
\begin{tikzpicture}[xscale=.8]
\node[SO] (A) at  (1,0)  {$m$};
\node[Sp] (B) at  (0,1.0)  {$\frac{m-3}{2}$};
\node[Sp] (D) at  (2,1.0)  {$\frac{m-3}{2}$};
\node[Sp] (F) at  (2,-1.0)  {$\frac{m-3}{2}$};
\node[Sp] (G) at  (0,-1.0)  {$\frac{m-3}{2}$};
\draw (B)--(A);
\draw (D)--(A);
\draw (F)--(A);
\draw (G)--(A);
\end{tikzpicture}\ , 
\begin{tikzpicture}[xscale=.8]
\node[SO] (A) at  (1,0)  {$m$};
\node[Sp] (B) at  (0,1.0)  {$\frac{m-3}{2}$};
\node[Sp] (D) at  (2,1.0)  {$\frac{m+1}{2}$};
\node[Sp] (F) at  (2,-1.0)  {$\frac{m-3}{2}$};
\node[Sp] (G) at  (0,-1.0)  {$\frac{m-3}{2}$};
\draw (B)--(A);
\draw (D)--(A);
\draw (F)--(A);
\draw (G)--(A);
\end{tikzpicture}\ .
\end{itemize}

All these graphs are of affine D type. Some of them can be viewed as degenerate cases of the graphs already counted. Some of them can be extended. Those containing $\SO(8)-\USp(6)$ are already counted, and so we do not repeat them. Clearly, in the above graphs, only the branches starting with $\SO(m)-\USp(\frac{m+4}{2})$, $\SO(m)-\USp(\frac{m}{2})$ and $\SO(m)-\USp(\frac{m+1}{2})$ have the potential to be extended. Pick one such potential branch. One can calculate the branch current in the branch as a function of $m$ and then, for a particular $m$, any branch having zero-th node $\SO(m)$ and the particular branch current would give one an allowed extended graph.  
We would like to comment that there are only a finite number of extensions possible. For instance, consider the branches starting with $\SO(m)-\USp(\frac{m+4}{2})$ for $m=4k$. For this case, the branch current, $I=\frac{m-8}{2}$, which implies that the third node in the branch is $\SO(n\leq8)$. As there are finite number of such vertices and the current in the branch is strictly positive, there are only finite number of possible extensions of this particular branch. Similar arguments for other branches prove our claim. All the extensions are non-Dynkin in nature.

\paragraph{3-valent:} First of all, the only case having an $\SO-\SU$ branch is:
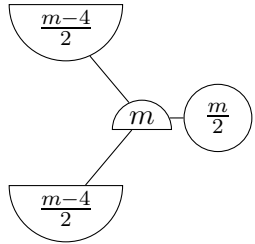
\[
\begin{tikzpicture}
\node[SO] (A) at  (1,0)  {$m$};
\node[SU] (B) at  (2,0)  {$\frac{m}{2}$};
\node[Sp] (P) at  (0,1.2)  {$\frac{m-4}{2}$};
\node[Sp] (Q) at  (0,-1.2)  {$\frac{m-4}{2}$};
\draw (A)--(B);
\draw (P)--(A);
\draw (Q)--(A);
\end{tikzpicture}
\] for $m=4k, k\geq2$. So, now we discuss only $\SO-\USp$ branches. Call the central node $\SO(m)$. Call the branch currents in three directions as $I_1$, $I_2$ and $I_3$. The branch $i$ could have length of $k$ nodes (including the zero-th node $\SO(m)$) only if $0\leq I_i\leq \frac{m-1+(-1)^k}{k}$. Also, we must have $I_1+I_2+I_3\geq m-2$. We can easily construct all the graphs having central node as $\SO(m)$ by taking three branches whose zero-th node is $\SO(m)$ and whose branch currents satisfy the above two inequalities. We now discuss some important properties of these graphs by dividing them into the following \emph{categories}:

\begin{itemize}
\item If the length of two branches is 2, then the length of the third branch is not restricted by any finite number for even $m$. When $m$ is even, and the current in the third branch is zero, we obtain finite and affine D type Dynkin graphs along with two exceptional non-Dynkin graphs. These all have been counted earlier when we discussed the case of $\SO-\USp$ trunk having more than one node. If the current in the third branch is not zero, all the graphs are of finite or affine D type. For odd $m$, the length of third branch cannot be more than $m-2$. All these graphs are of finite or affine D type. There can be various decorations in the branches.
\item Let us  consider the case when the length of first branch is 2 and the length of second branch is 3. 
	\begin{itemize}
	\item
For $k_3=10$, combining the above two inequalities, we obtain that $m\leq20$. But for the USp labellings to be an even number, $I_i$ for all $i$ should be even or odd respectively for even or odd $m$. This implies that we can't have even $m$ because even for $m=20$, the maximum labelling of the $\USp$ vertex at the end of the branch is at most 0. If $I_3=1$, then requiring that the labelling on last $\USp$ vertex $\geq2$ implies that $13\leq m\leq19$. We would like to note here the general fact that if a $\SO-\USp$ branch has zero-th node $\SO(m=4k+l)$ where $1\leq l\leq4$, then the branch current, $I\leq (m-4+l)/2$. Because $\frac{m-4+l}{2}\leq\frac{m}{2}$, we quoted earlier that for a branch to have length 2, the branch current must satisfy $I\leq m/2$. So, we see that the first inequality is modified for $I_1$. Combining modified first inequality and second inequality for $I_3=1$, we obtain $\frac{m-4+l}{2}+\frac{m-2}{3}\geq I_1+I_2\geq m-3$ which implies that $4\geq l\geq\frac{m-2}{3}$ which means that $m=13$ and $l=4$, which is a contradiction. Hence, there doesn't exist any graph in this category having $k_3\geq10$. 
	\item
For $k_3=9$, combining the above two inequalities, we obtain that $m\leq20$. Requiring that the the labelling on last $\SO$ vertex $\geq3$, we find that $I_3=1,2$. We also find that for $I_3=1$, $11\leq m$ and for $I_3=2$, $m=20$ for which both the above inequalities saturate in the sense that $I_1+I_2+I_3$ is sandwiched between the two inequalities leaving a unique solution in this case. Now, for $I_3=1$, we have already shown that all the $m\geq13$ aren't allowed, which means that the only potential possibility for odd $m$ is $m=11$. And indeed, the modified first inequality and second inequality saturate for $m=11$ leaving a unique solution in this case. Hence, the graphs in this category are:
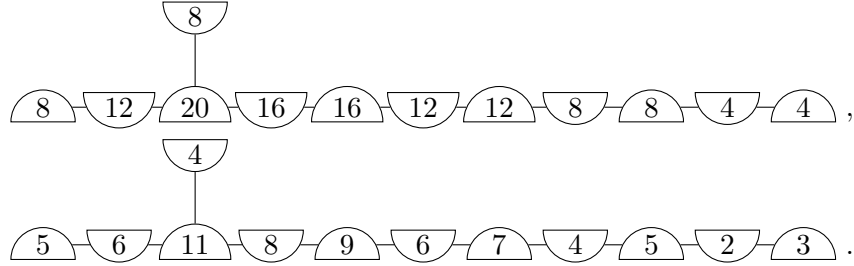
\begin{gather*}
\begin{tikzpicture}
\node[SO] (A) at  (1,0)  {8};
\node[Sp] (B) at  (2,0)  {12};
\node[SO] (C) at  (3,0)  {20};
\node[Sp] (D) at  (4,0)  {16};
\node[SO] (E) at  (5,0)  {16};
\node[Sp] (F) at  (6,0)  {12};
\node[SO] (G) at  (7,0)  {12};
\node[Sp] (H) at  (8,0)  {8};
\node[SO] (I) at  (9,0)  {8};
\node[Sp] (J) at  (10,0)  {4};
\node[SO] (K) at  (11,0)  {4};
\node[Sp] (P) at  (3,1.2)  {8};
\draw (A)--(B)--(C)--(D)--(E)--(F)--(G)--(H)--(I)--(J)--(K);
\draw (P)--(C);
\end{tikzpicture}\ ,
\\
\begin{tikzpicture}
\node[SO] (A) at  (1,0)  {5};
\node[Sp] (B) at  (2,0)  {6};
\node[SO] (C) at  (3,0)  {11};
\node[Sp] (D) at  (4,0)  {8};
\node[SO] (E) at  (5,0)  {9};
\node[Sp] (F) at  (6,0)  {6};
\node[SO] (G) at  (7,0)  {7};
\node[Sp] (H) at  (8,0)  {4};
\node[SO] (I) at  (9,0)  {5};
\node[Sp] (J) at  (10,0)  {2};
\node[SO] (K) at  (11,0)  {3};
\node[Sp] (P) at  (3,1.2)  {4};
\draw (A)--(B)--(C)--(D)--(E)--(F)--(G)--(H)--(I)--(J)--(K);
\draw (P)--(C);
\end{tikzpicture}\ .
\end{gather*}
Both of these graphs are non-Dynkin. There can be various decorations, wherever possible.
	\item
Similarly, one can construct allowed graphs for $k_3=7,8$. The maximum values of central node are 44 and 32 for $k_3=7$ and $k_3=8$ respectively. This implies that these graphs are finite in number. There can also be further branchings in this case along with the decorations. These all are non-Dynkin in nature.
	\item 
	For $k_3=6$, we obtain that there isn't any restriction placed on $m$ by the combination of both the inequalities. This hints at the possibility that for $k_3\leq6$ we can take the labellings in the graph to be arbitrarily large such that various relations between various labellings are obeyed. Indeed, this is the case. For instance, one can consider, for $m=12k+8$:

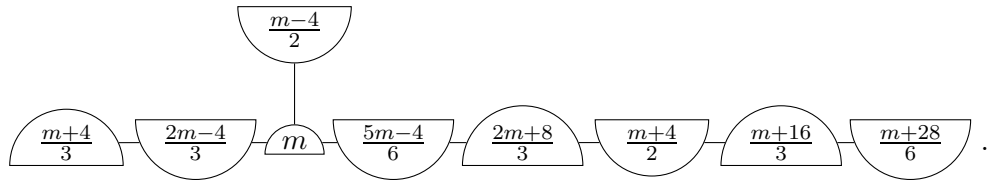
\begin{tikzpicture}
\node[SO] (A) at  (0,0)  {$\frac{m+4}{3}$};
\node[Sp] (B) at  (1.7,0)  {$\frac{2m-4}{3}$};
\node[SO] (C) at  (3,0)  {$m$};
\node[Sp] (D) at  (4.3,0)  {$\frac{5m-4}{6}$};
\node[SO] (E) at  (6,0)  {$\frac{2m+8}{3}$};
\node[Sp] (F) at  (7.7,0)  {$\frac{m+4}{2}$};
\node[SO] (G) at  (9.4,0)  {$\frac{m+16}{3}$};
\node[Sp] (H) at  (11.1,0)  {$\frac{m+28}{6}$};
\node[Sp] (P) at  (3,1.5)  {$\frac{m-4}{2}$};
\draw (A)--(B)--(C)--(D)--(E)--(F)--(G)--(H);
\draw (P)--(C);
\end{tikzpicture}\ .

Therefore, in this category, we have finite E type Dynkin graphs and affine E$_8$. The special property that these Dynkin graphs enjoy is that there is no limit to the size of labellings at any vertex. We can have various decorations, wherever possible. 

There can't be further branching in the second branch because the only available node for further branching in this direction is of $\USp$ type which doesn't support further branching. For further branching in the third direction, either $I_3=1$ or $I_3=2$. If $I_3=2$, then the current must remain 2 throughout the branch and $m=4k$. The combination of first and second inequality tells us that $m\leq20$ for further branching to occur. Thus, we have:
\begin{itemize}
\item For $m=20$, 
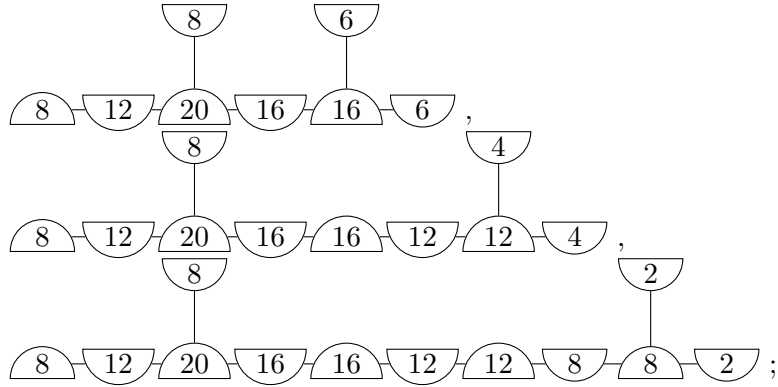
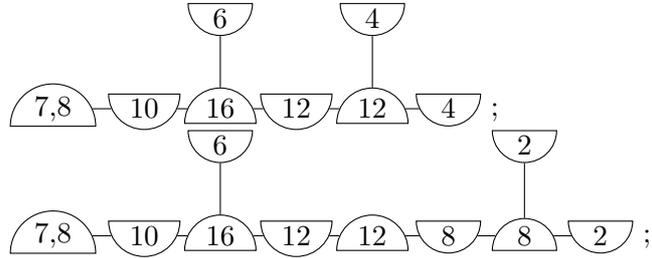
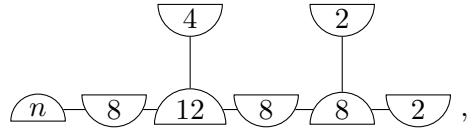
\[\begin{array}{l}
\begin{tikzpicture}
\node[SO] (A) at  (1,0)  {8};
\node[Sp] (B) at  (2,0)  {12};
\node[SO] (C) at  (3,0)  {20};
\node[Sp] (D) at  (4,0)  {16};
\node[SO] (E) at  (5,0)  {16};
\node[Sp] (F) at  (6,0)  {6};
\node[Sp] (P) at  (3,1.2)  {8};
\node[Sp] (Q) at  (5,1.2)  {6};
\draw (A)--(B)--(C)--(D)--(E)--(F);
\draw (Q)--(E);
\draw (P)--(C);
\end{tikzpicture}\ ,\\
\begin{tikzpicture}
\node[SO] (A) at  (1,0)  {8};
\node[Sp] (B) at  (2,0)  {12};
\node[SO] (C) at  (3,0)  {20};
\node[Sp] (D) at  (4,0)  {16};
\node[SO] (E) at  (5,0)  {16};
\node[Sp] (F) at  (6,0)  {12};
\node[SO] (G) at  (7,0)  {12};
\node[Sp] (H) at  (8,0)  {4};
\node[Sp] (P) at  (3,1.2)  {8};
\node[Sp] (Q) at  (7,1.2)  {4};
\draw (A)--(B)--(C)--(D)--(E)--(F)--(G)--(H);
\draw (Q)--(G);
\draw (P)--(C);
\end{tikzpicture}\ ,\\
\begin{tikzpicture}
\node[SO] (A) at  (1,0)  {8};
\node[Sp] (B) at  (2,0)  {12};
\node[SO] (C) at  (3,0)  {20};
\node[Sp] (D) at  (4,0)  {16};
\node[SO] (E) at  (5,0)  {16};
\node[Sp] (F) at  (6,0)  {12};
\node[SO] (G) at  (7,0)  {12};
\node[Sp] (H) at  (8,0)  {8};
\node[SO] (I) at  (9,0)  {8};
\node[Sp] (J) at  (10,0)  {2};
\node[Sp] (P) at  (3,1.2)  {8};
\node[Sp] (Q) at  (9,1.2)  {2};
\draw (A)--(B)--(C)--(D)--(E)--(F)--(G)--(H)--(I)--(J);
\draw (Q)--(I);
\draw (P)--(C);
\end{tikzpicture}\ ;
\end{array}\]
\item For $m=16$, 
\[\begin{array}{l}
\begin{tikzpicture}
\node[SO] (A) at  (0.8,0)  {7,8};
\node[Sp] (B) at  (2,0)  {10};
\node[SO] (C) at  (3,0)  {16};
\node[Sp] (D) at  (4,0)  {12};
\node[SO] (E) at  (5,0)  {12};
\node[Sp] (F) at  (6,0)  {4};
\node[Sp] (P) at  (3,1.2)  {6};
\node[Sp] (Q) at  (5,1.2)  {4};
\draw (A)--(B)--(C)--(D)--(E)--(F);
\draw (Q)--(E);
\draw (P)--(C);
\end{tikzpicture}\ ;
\\
\begin{tikzpicture}
\node[SO] (A) at  (0.8,0)  {7,8};
\node[Sp] (B) at  (2,0)  {10};
\node[SO] (C) at  (3,0)  {16};
\node[Sp] (D) at  (4,0)  {12};
\node[SO] (E) at  (5,0)  {12};
\node[Sp] (F) at  (6,0)  {8};
\node[SO] (G) at  (7,0)  {8};
\node[Sp] (H) at  (8,0)  {2};
\node[Sp] (P) at  (3,1.2)  {6};
\node[Sp] (Q) at  (7,1.2)  {2};
\draw (A)--(B)--(C)--(D)--(E)--(F)--(G)--(H);
\draw (Q)--(G);
\draw (P)--(C);
\end{tikzpicture}\ ;
\end{array}\]
\item For $m=12$, \[
\begin{tikzpicture}
\node[SO] (A) at  (1,0)  {$n$};
\node[Sp] (B) at  (2,0)  {8};
\node[SO] (C) at  (3,0)  {12};
\node[Sp] (D) at  (4,0)  {8};
\node[SO] (E) at  (5,0)  {8};
\node[Sp] (F) at  (6,0)  {2};
\node[Sp] (P) at  (3,1.2)  {4};
\node[Sp] (Q) at  (5,1.2)  {2};
\draw (A)--(B)--(C)--(D)--(E)--(F);
\draw (Q)--(E);
\draw (P)--(C);
\end{tikzpicture}\ , 
\] where $6\leq n\leq8$. 

If $I_3=1$, then the modified combination of first and second inequalities tells us that $m\leq2+3l$. If $l=1$, then $m=5$ which can't have a branch of length 3. If $l=3$, then $m=7,11$. For $m=7$, we only get a degenerate case which has been counted earlier. For $m=11$, we obtain:
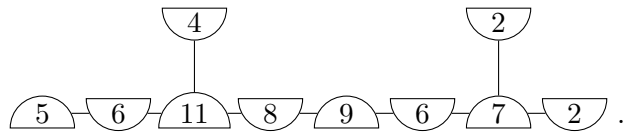
\[
\begin{tikzpicture}
\node[SO] (A) at  (1,0)  {5};
\node[Sp] (B) at  (2,0)  {6};
\node[SO] (C) at  (3,0)  {11};
\node[Sp] (D) at  (4,0)  {8};
\node[SO] (E) at  (5,0)  {9};
\node[Sp] (F) at  (6,0)  {6};
\node[SO] (G) at  (7,0)  {7};
\node[Sp] (H) at  (8,0)  {2};
\node[Sp] (P) at  (3,1.2)  {4};
\node[Sp] (Q) at  (7,1.2)  {2};
\draw (A)--(B)--(C)--(D)--(E)--(F)--(G)--(H);
\draw (P)--(C);
\draw (Q)--(G);
\end{tikzpicture}\ . 
\]
\end{itemize}
All of the further branchings in this category render the graphs non-Dynkin in nature.
	\end{itemize}
\item Let us  consider the case when the length of first branch is 2 and the length of second branch is 4. 
	\begin{itemize}
	\item For $k_3=7$, combining modified first inequality with second, we obtain that $m\leq\frac{14l-8}{3}$. Clearly for $l=1,2$ and for the allowed values of $m$ corresponding to these $l$, there isn't any branch having length 7. For $l=3$, $m=7$ doesn't have any branch of length 7, $m=11$ must have potential $I_3=1$ as all higher $I_3$ render the branch lengths too small and, so, using the fact $I_3=1$ in the combination of modified first and second inequalities, we obtain $m=11\geq12$ which is a contradiction. For $l=4$, $I_3$ must be 2 and only for $m=16$ does this ensure that the last $\USp$ label is positive. As the inequalities are saturated in this case, the only allowed graph in this category having $k_3=7$ is:
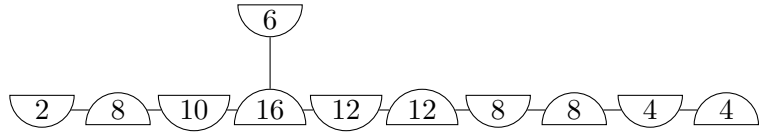
\[
\begin{tikzpicture}
\node[Sp] (X) at  (0,0)  {2};
\node[SO] (A) at  (1,0)  {8};
\node[Sp] (B) at  (2,0)  {10};
\node[SO] (C) at  (3,0)  {16};
\node[Sp] (D) at  (4,0)  {12};
\node[SO] (E) at  (5,0)  {12};
\node[Sp] (F) at  (6,0)  {8};
\node[SO] (G) at  (7,0)  {8};
\node[Sp] (H) at  (8,0)  {4};
\node[SO] (I) at  (9,0)  {4};
\node[Sp] (P) at  (3,1.2)  {6};
\draw (X)--(A)--(B)--(C)--(D)--(E)--(F)--(G)--(H)--(I);
\draw (P)--(C);
\end{tikzpicture}
\]
which is a non-Dynkin graph. As the third branch of this graph is not extendable, there are no graphs in this category having $k_3\geq8$.
	\item Similarly, one can find allowed graphs for $k_3=5,6$. The maximum values of central node are 32 and 24 for $k_3=5$ and $k_3=6$ respectively. This implies that these graphs are finite in number. There can also be further branchings in this case along with the decorations. These all are non-Dynkin in nature.

	\item For $k_3=4$, we obtain that there isn't any restriction placed by both the inequalities. This hints at the possibility that for $k_3=4$ we can take the labellings in the graph to be arbitrarily large such that various relations between various labellings are obeyed. Indeed, this is the case. For instance, one can consider, for $m=8k$:
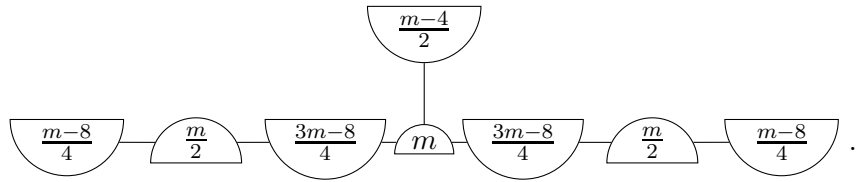
\[
\begin{tikzpicture}
\node[Sp] (A) at  (0,0)  {$\frac{m-8}{4}$};
\node[SO] (B) at  (1.7,0)  {$\frac{m}{2}$};
\node[Sp] (C) at  (3.4,0)  {$\frac{3m-8}{4}$};
\node[SO] (D) at  (4.7,0)  {$m$};
\node[Sp] (E) at  (6,0)  {$\frac{3m-8}{4}$};
\node[SO] (F) at  (7.7,0)  {$\frac{m}{2}$};
\node[Sp] (G) at  (9.4,0)  {$\frac{m-8}{4}$};
\node[Sp] (P) at  (4.7,1.5)  {$\frac{m-4}{2}$};
\draw (A)--(B)--(C)--(D)--(E)--(F)--(G);
\draw (P)--(D);
\end{tikzpicture}\ .
\]
Therefore, in this category, we have affine E$_7$. The special property that these Dynkin graphs enjoy is that there is no limit to the size of labellings at any vertex. We can have various decorations, wherever possible.

For further branching in third branch, either $I_3=1$ or $I_3=2$. If $I_3=2$, then the current must remain 2 throughout the branch and $m=4k$. The combination of first and second inequality tells us that $m\leq16$ for further branching to occur. Thus, we have:
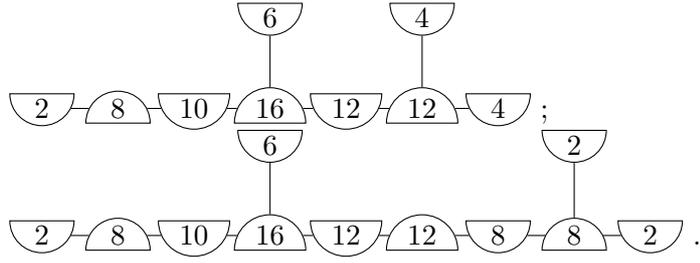
\[\begin{array}{l}
\begin{tikzpicture}
\node[Sp] (X) at  (0,0)  {2};
\node[SO] (A) at  (1,0)  {8};
\node[Sp] (B) at  (2,0)  {10};
\node[SO] (C) at  (3,0)  {16};
\node[Sp] (D) at  (4,0)  {12};
\node[SO] (E) at  (5,0)  {12};
\node[Sp] (F) at  (6,0)  {4};
\node[Sp] (P) at  (3,1.2)  {6};
\node[Sp] (Q) at  (5,1.2)  {4};
\draw (X)--(A)--(B)--(C)--(D)--(E)--(F);
\draw (Q)--(E);
\draw (P)--(C);
\end{tikzpicture}\ ;
\\
\begin{tikzpicture}
\node[Sp] (X) at  (0,0)  {2};
\node[SO] (A) at  (1,0)  {8};
\node[Sp] (B) at  (2,0)  {10};
\node[SO] (C) at  (3,0)  {16};
\node[Sp] (D) at  (4,0)  {12};
\node[SO] (E) at  (5,0)  {12};
\node[Sp] (F) at  (6,0)  {8};
\node[SO] (G) at  (7,0)  {8};
\node[Sp] (H) at  (8,0)  {2};
\node[Sp] (P) at  (3,1.2)  {6};
\node[Sp] (Q) at  (7,1.2)  {2};
\draw (X)--(A)--(B)--(C)--(D)--(E)--(F)--(G)--(H);
\draw (Q)--(G);
\draw (P)--(C);
\end{tikzpicture}\ . 
\end{array}\]
If $I_3=1$, then the modified combination of first and second inequalities tells us that $m\leq2(l+2)$. If $l=1$, then $m=5$ which can't even have a branch of length 3. If $l=3$, then $m=7$. For $m=7$, we only get a degenerate case which has been counted earlier.

For further branching in the second branch either $I_2=1$ or $I_2=2$. If $I_2=2$, then the current must remain 2 throughout the branch and $m=4k$. Combining the first and second inequalities, we obtain that, for $I_2=2$, $I_3\geq\frac{m-8}{2}$. Combining this again with first inequality for $I_3$, we see that only possibility is $m=12$ with $k_3=4,5$ and the graphs are:
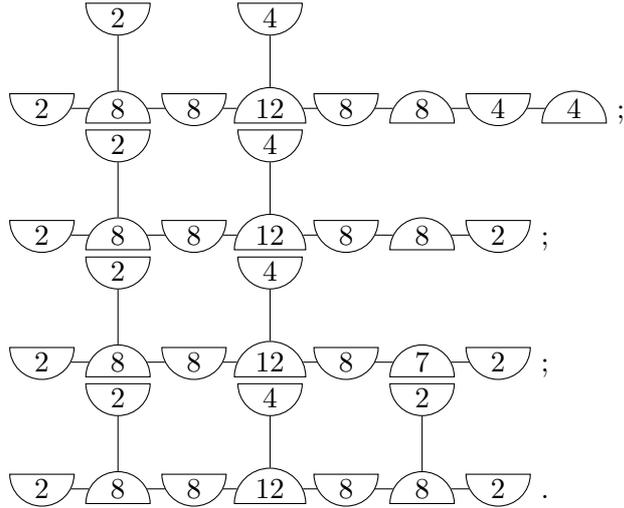
\[\begin{array}{l}
\begin{tikzpicture}
\node[Sp] (X) at  (0,0)  {2};
\node[Sp] (Y) at  (1,1.2)  {2};
\node[SO] (A) at  (1,0)  {8};
\node[Sp] (B) at  (2,0)  {8};
\node[SO] (C) at  (3,0)  {12};
\node[Sp] (D) at  (4,0)  {8};
\node[SO] (E) at  (5,0)  {8};
\node[Sp] (F) at  (6,0)  {4};
\node[SO] (G) at  (7,0)  {4};
\node[Sp] (P) at  (3,1.2)  {4};
\draw (X)--(A)--(B)--(C)--(D)--(E)--(F)--(G);
\draw (Y)--(A);
\draw (P)--(C);
\end{tikzpicture}\ ;
\\
\begin{tikzpicture}
\node[Sp] (X) at  (0,0)  {2};
\node[Sp] (Y) at  (1,1.2)  {2};
\node[SO] (A) at  (1,0)  {8};
\node[Sp] (B) at  (2,0)  {8};
\node[SO] (C) at  (3,0)  {12};
\node[Sp] (D) at  (4,0)  {8};
\node[SO] (E) at  (5,0)  {8};
\node[Sp] (F) at  (6,0)  {2};
\node[Sp] (P) at  (3,1.2)  {4};
\draw (X)--(A)--(B)--(C)--(D)--(E)--(F);
\draw (Y)--(A);
\draw (P)--(C);
\end{tikzpicture}\ ;
\\
\begin{tikzpicture}
\node[Sp] (X) at  (0,0)  {2};
\node[Sp] (Y) at  (1,1.2)  {2};
\node[SO] (A) at  (1,0)  {8};
\node[Sp] (B) at  (2,0)  {8};
\node[SO] (C) at  (3,0)  {12};
\node[Sp] (D) at  (4,0)  {8};
\node[SO] (E) at  (5,0)  {7};
\node[Sp] (F) at  (6,0)  {2};
\node[Sp] (P) at  (3,1.2)  {4};
\draw (X)--(A)--(B)--(C)--(D)--(E)--(F);
\draw (Y)--(A);
\draw (P)--(C);
\end{tikzpicture}\ ;
\\
\begin{tikzpicture}
\node[Sp] (X) at  (0,0)  {2};
\node[Sp] (Y) at  (1,1.2)  {2};
\node[SO] (A) at  (1,0)  {8};
\node[Sp] (B) at  (2,0)  {8};
\node[SO] (C) at  (3,0)  {12};
\node[Sp] (D) at  (4,0)  {8};
\node[SO] (E) at  (5,0)  {8};
\node[Sp] (F) at  (6,0)  {2};
\node[Sp] (P) at  (3,1.2)  {4};
\node[Sp] (Q) at  (5,1.2)  {2};
\draw (X)--(A)--(B)--(C)--(D)--(E)--(F);
\draw (Y)--(A);
\draw (Q)--(E);
\draw (P)--(C);
\end{tikzpicture}\ .
\end{array}\]
 For $I_2=1$, we obtain $I_3\geq\frac{m-l-2}{2}$ and combining the first inequality for $I_3$ we see that it isn't possible to have any new graphs. 

Hence, all of the further branchings in this category render the graphs non-Dynkin in nature.
	\end{itemize}
	
\item Let us  consider the case when the length of first branch is 2 and the length of second branch is 5. For $k_3=5$, combing the modified first inequality with second, we obtain that $m\leq5l-8$. Clearly, there are no graphs for $l=1,2$. For $l=3$, $I_3=1$ and for $l=4$, $I_3=2$. These imply that there is only a single possible graph for both $l=3,4$:
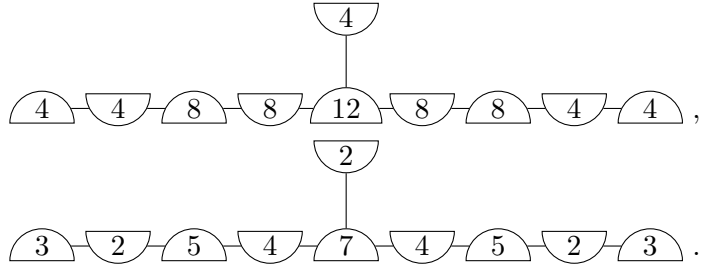
\begin{gather*}
\begin{tikzpicture}
\node[SO] (Y) at  (-1,0)  {4};
\node[Sp] (X) at  (0,0)  {4};
\node[SO] (A) at  (1,0)  {8};
\node[Sp] (B) at  (2,0)  {8};
\node[SO] (C) at  (3,0)  {12};
\node[Sp] (D) at  (4,0)  {8};
\node[SO] (E) at  (5,0)  {8};
\node[Sp] (F) at  (6,0)  {4};
\node[SO] (G) at  (7,0)  {4};
\node[Sp] (P) at  (3,1.2)  {4};
\draw (Y)--(X)--(A)--(B)--(C)--(D)--(E)--(F)--(G);
\draw (P)--(C);
\end{tikzpicture}\ ,
\\
\begin{tikzpicture}
\node[SO] (Y) at  (-1,0)  {3};
\node[Sp] (X) at  (0,0)  {2};
\node[SO] (A) at  (1,0)  {5};
\node[Sp] (B) at  (2,0)  {4};
\node[SO] (C) at  (3,0)  {7};
\node[Sp] (D) at  (4,0)  {4};
\node[SO] (E) at  (5,0)  {5};
\node[Sp] (F) at  (6,0)  {2};
\node[SO] (G) at  (7,0)  {3};
\node[Sp] (P) at  (3,1.2)  {2};
\draw (Y)--(X)--(A)--(B)--(C)--(D)--(E)--(F)--(G);
\draw (P)--(C);
\end{tikzpicture}\ .
\end{gather*}
These are both non-Dynkin graphs and because the third branch in these graphs can't be extended, there exist no graphs in this category having $k_3\geq6$.
\item Let us  consider the case when the length of first branch is 3 and the length of second branch is 3. For $k_3=3$, we see that both the inequalities are saturated irrespective of the value of $m$. But, every $m$ isn't allowed because the current $I=\frac{m-2}{3}$ need not be integer for every $m$ or the labelling on $\USp$ vertex need not be even. Keeping these in mind we find that all the graphs in this category having $k_3=3$ are:
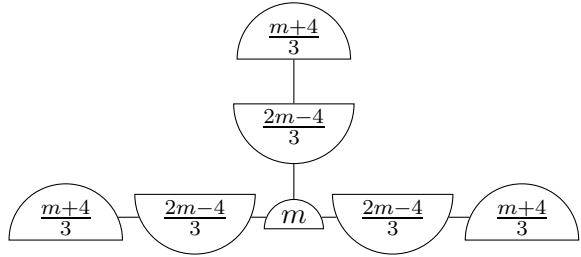
\[
\begin{tikzpicture}
\node[SO] (A) at  (0,0)  {$\frac{m+4}{3}$};
\node[Sp] (B) at  (1.7,0)  {$\frac{2m-4}{3}$};
\node[SO] (C) at  (3,0)  {$m$};
\node[Sp] (D) at  (4.3,0)  {$\frac{2m-4}{3}$};
\node[SO] (E) at  (6,0)  {$\frac{m+4}{3}$};
\node[Sp] (P) at  (3,1.2)  {$\frac{2m-4}{3}$};
\node[SO] (Q) at  (3,2.4)  {$\frac{m+4}{3}$};
\draw (A)--(B)--(C)--(D)--(E);
\draw (Q)--(P)--(C);
\end{tikzpicture} 
\] 
where $m=3k+2$ and the graph is of affine E$_6$ type. As this graph is not extendable, there doesn't exist any graph in this category having $k_3\geq4$ and also there are no other categories to consider. There can be various decorations, wherever possible.
\end{itemize}

\paragraph{2-valent:} To remove the graphs which have already been counted before, we may impose that the usual currents in both the directions are strictly positive. We have the following two possibilities:

\begin{itemize}
\item We can take an $\SO(m)-\SU(n_1)$ branch along with an $\SO(m)-\USp(n_2)$ with branch currents $I_1$ and $I_2$ respectively, such that $I_1+I_2\geq0$. All such configurations give all the allowed graphs in this category. We can have further branching along the $\SO-\USp$ direction if $I_2=2$, which means that $n_1=m/2$ and hence the length of the $\SO-\SU$ branch is only 2. All such graphs are of finite A or D type.
\item We can take two $\SO-\USp$ branches. Then, all such graphs can be viewed as a degenerate case of the $\SO-\USp$ trunk with more than one node, and can be obtained by shrinking the trunk to exactly one $\SO$ node. All these graphs are of finite A, finite D or affine D type.
\end{itemize}
We can have various decorations, wherever possible.

We also have cases involving large branches of length-1, 
listed in Sec.~\ref{what?}.
The two possibilities are \[
\begin{tikzpicture}
\node[1gon] (A) at  (1,0)  {$\spinor$};
\node[SO] (B) at  (2,0)  {14};
\node[Sp] (C) at  (3,0)  {8};
\node[SO] (D) at  (4,0)  {6};
\draw (A)--(B)--(C)--(D);
\end{tikzpicture}\ , \qquad
\begin{tikzpicture}
\node[1gon] (A) at  (1,0)  {$\spinor$};
\node[SO] (B) at  (2,0)  {14};
\node[Sp] (C) at  (3,0)  {6};
\draw (A)--(B)--(C);
\end{tikzpicture}\ .
\] 

\paragraph{1-valent:} These are precisely the branches with usual current $I>0$.

\paragraph{Comment:} Notice that the only graphs in which we could take the labelling of every vertex to be arbitrarily large were found to be of finite and affine Dynkin type. The labellings on all the non-Dynkin graphs are bounded and such graphs are finite in number.

\subsubsection{$\USp$ trunk with exactly one node}\label{usp-trunk}

\paragraph{4-valent or higher:} There are no such graphs.

\paragraph{3-valent:} First of all, the only cases having a $\USp-\SU$ branch for general $n$  are: \[
\begin{tikzpicture}
\node[SO] (A) at  (0,0)  {$\frac{n+4}{2}$};
\node[Sp] (B) at  (1.3,0)  {$n$};
\node[SO] (C) at  (2.6,0)  {$\frac{n+4}{2}$};
\node[SU] (P) at  (1.3,1.2)  {$\frac{n}{2}$};
\draw (A)--(B)--(C);
\draw (P)--(B);
\end{tikzpicture}\ .
\]
For $n=4$, things are a little special, as three types of branches \[
\begin{tikzpicture}
\node[Sp] (A) at (0,0) {$4$};
\node[SO] (B) at (1,0) {$4$};
\draw (A)--(B);
\end{tikzpicture}\ , 
\quad
\begin{tikzpicture}
\node[Sp] (A) at (0,0) {$4$};
\node[SU] (B) at (1,0) {$2$};
\draw (A)--(B);
\end{tikzpicture}\ , 
\quad
\begin{tikzpicture}
\node[Sp] (A) at (0,0) {$4$};
\node[SO] (B) at (0.5,0) {$5$};
\node[Sp] (C) at (1.5,0) {$2$};
\node[SO] (D) at (2.5,0) {$3$};
\draw  (B)--(C)--(D);
\end{tikzpicture} 
\] have exactly the same branch current $2$. 
Therefore we can take three branches out of these three types of branches, allowing repetition, to have a superconformal gauge theory.

So, now we discuss only $\USp-\SO$ branches. Call the central node $\USp(n)$. Call the branch currents in three directions as $I_1$, $I_2$ and $I_3$. The branch $i$ could have length of $k$ nodes (including the zero-th node $\USp(n)$) only if $0\leq I_i\leq \frac{n+1+(-1)^{k+1}}{k}$. Also, we must have $I_1+I_2+I_3\geq n+2$. We can easily construct all the graphs having central node as $\USp(n)$ by taking three branches whose zero-th node is $\USp(n)$ and whose branch currents satisfy the above two inequalities. We now discuss some important properties of these graphs by dividing them into the following \emph{categories}:
\begin{itemize}
\item Let us  consider the case when $k_1=k_2=2$. Even if $I_1=\frac{n}{2}$ and $I_2=\frac{n}{2}$, the second inequality tells us that $I_3\geq2$. This means that the currents always flow out of a trivalent $\USp$ vertex. Hence, the length of third branch is bounded by $\frac{n}{2}$. When $I_3=2$ and $n=4k$, the third branch can further branch out at an $\SO$ vertex. Hence, the graphs in this category are of finite and affine D type. There can be various decorations too.
\item Let us  consider the case when $k_1=2$ and $k_2=3$. For $k_3=6$, the combination of above two inequalities tells us that $2\geq6$, which is a contradiction. Hence, there are no graphs in this case having $k_3\geq6$. For $k_3=5$, the combination tells us that $n\geq28$. Again, it can be easily seen that there do exist graphs of this type and the labelling at any vertex of the graphs having $k_3\leq5$ can be taken to be arbitrarily large. For instance, consider for $n=6k+4$:
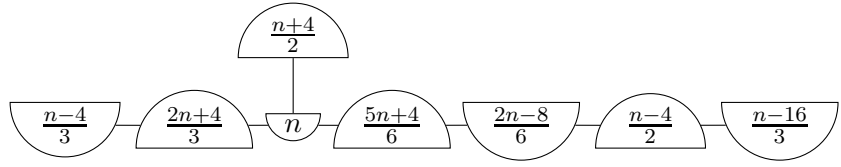
\[
\begin{tikzpicture}
\node[Sp] (A) at  (0,0)  {$\frac{n-4}{3}$};
\node[SO] (B) at  (1.7,0)  {$\frac{2n+4}{3}$};
\node[Sp] (C) at  (3,0)  {$n$};
\node[SO] (D) at  (4.3,0)  {$\frac{5n+4}{6}$};
\node[Sp] (E) at  (6,0)  {$\frac{2n-8}{6}$};
\node[SO] (F) at  (7.7,0)  {$\frac{n-4}{2}$};
\node[Sp] (G) at  (9.4,0)  {$\frac{n-16}{3}$};
\node[SO] (P) at  (3,1.2)  {$\frac{n+4}{2}$};
\draw (A)--(B)--(C)--(D)--(E)--(F)--(G);
\draw (P)--(C);
\end{tikzpicture}
\]
Hence, there are only finite E type Dynkin graphs in this category. There can't be any further branching in the third branch as for it to occur, $I_3$ must be 2 but combining this fact with the above two inequalities we find that for this to happen $n\leq4$, which is too small. Now, while discussing further branching in second branch, we can impose $k_3=3$ as for all higher $k_3$ we can always cut the graph up to $k_3=3$ and this subgraph remains allowed. Now, we see, by the same argument as above, that it is impossible to have further branching in second branch too. There still can be decorations which can occur if branch currents are more than 2.
\item Let us  consider the case when $k_1=2$ and $k_2=4$. To avoid double counting of some graphs of previous category, we impose that the graphs of this category must have $k_3\geq4$. For $k_3=4$, we find from the above two inequalities that $n\geq n+2$, which is a contradiction. Hence, there are no graphs in this category and therefore in other categories having $k_2>4$.
\item Let us  consider the case when $k_1=3$ and $k_2=3$. Then, for $k_3=3$, the above two inequalities are saturated and hence the only solution is:
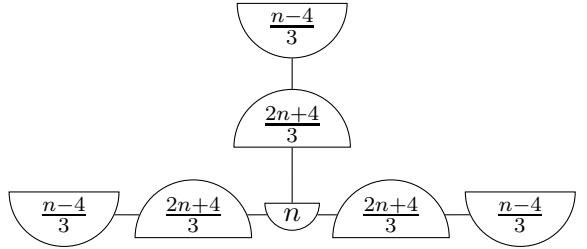
\[
\begin{tikzpicture}
\node[Sp] (A) at  (0,0)  {$\frac{n-4}{3}$};
\node[SO] (B) at  (1.7,0)  {$\frac{2n+4}{3}$};
\node[Sp] (C) at  (3,0)  {$n$};
\node[SO] (D) at  (4.3,0)  {$\frac{2n+4}{3}$};
\node[Sp] (E) at  (6,0)  {$\frac{n-4}{3}$};
\node[SO] (P) at  (3,1.2)  {$\frac{2n+4}{3}$};
\node[Sp] (Q) at  (3,2.5)  {$\frac{n-4}{3}$};
\draw (A)--(B)--(C)--(D)--(E);
\draw (Q)--(P)--(C);
\end{tikzpicture}\ , 
\] 
where $n=6k+4$. This graph is of affine E$_6$ type. As this graph is non-extendable, there are no other 3-valent graphs left.
\end{itemize}

\paragraph{2-valent:} We can divide the analysis into the following categories:
\begin{itemize}
\item If there are two $\USp-\USp$ branches, then the only possibility is \[
\begin{tikzpicture}
\node[Sp] (A) at  (0,0)  {4};
\node[Sp] (B) at  (1,0)  {6};
\node[Sp] (C) at  (2,0)  {4};
\draw (A) --(B)--(C);
\end{tikzpicture}\ .
\]

\item If there's a $\USp-\USp$ and a $\USp-\SU$ branch, we have \[
\begin{tikzpicture}
\node[SU] (A) at  (1,0)  {2};
\node[Sp] (B) at  (2,0)  {4};
\node[Sp] (C) at  (3,0)  {4};
\draw (A) --(B)--(C);
\end{tikzpicture}\ , \quad
\begin{tikzpicture}
\node[SU] (A) at  (1,0)  {4};
\node[Sp] (B) at  (2,0)  {8};
\node[Sp] (C) at  (3,0)  {6};
\draw (A) --(B)--(C);
\end{tikzpicture}\ , \quad
\begin{tikzpicture}
\node[SU] (A) at  (1,0)  {3};
\node[Sp] (B) at  (2,0)  {6};
\node[Sp] (C) at  (3,0)  {4};
\draw (A) --(B)--(C);
\end{tikzpicture}\ , \quad
\begin{tikzpicture}
\node[SU] (Z) at  (0,0)  {2};
\node[SU] (A) at  (1,0)  {4};
\node[Sp] (B) at  (2,0)  {6};
\node[Sp] (C) at  (3,0)  {4};
\draw (Z)--(A) --(B)--(C);
\end{tikzpicture}\ .
\]

\item If there's a $\USp-\USp$ and a $\USp-\SO$ branch, we have 
\[
\begin{array}{rl}
\begin{tikzpicture}
\node[SO] (A) at  (1,0)  {4};
\node[Sp] (B) at  (2,0)  {4};
\node[Sp] (C) at  (3,0)  {4};
\draw	(A)--(B)--(C);
\end{tikzpicture}\ ; \\
\begin{tikzpicture}
\node[Sp] (A) at  (1,0)  {$m-2$};
\node[Sp] (B) at  (2.5,0)  {$m$};
\node[SO] (C) at  (3.5,0)  {8};
\draw	(A)--(B)--(C);
\end{tikzpicture}\ , & m=6,8,10,12; \\
\begin{tikzpicture}
\node[Sp] (A) at  (1,0)  {$m-2$};
\node[Sp] (B) at  (2.5,0)  {$m$};
\node[SO] (C) at  (3.5,0)  {7};
\draw	(A)--(B)--(C);
\end{tikzpicture}\ , & m=6,8,10; \\
\begin{tikzpicture}
\node[Sp] (A) at  (1,0)  {$m-2$};
\node[Sp] (B) at  (2.5,0)  {$m$};
\node[SO] (C) at  (3.5,0)  {6};
\draw	(A)--(B)--(C);
\end{tikzpicture}\ , & m=6,8; \\
\begin{tikzpicture}
\node[Sp] (A) at  (1.5,0)  {4};
\node[Sp] (B) at  (2.5,0)  {6};
\node[SO] (C) at  (3.5,0)  {5};
\draw	(A)--(B)--(C);
\end{tikzpicture}\ .
\end{array}
\]

Some of the $\USp-\SO$ branches above can be extended by replacing the branch by some other $\USp-\SO$ branch having the same $\USp$ vertex and same branch current. We include all such graphs here. Some of these graphs have further branching and decorations too. These are all of finite A or D type along with two exceptions arising out of the combination of two exceptional type of branches at an $\SO(8)$ node in a $\USp(6)-\SO(8)$ trunk above.

\item If there are two $\USp-\SU$ branches:

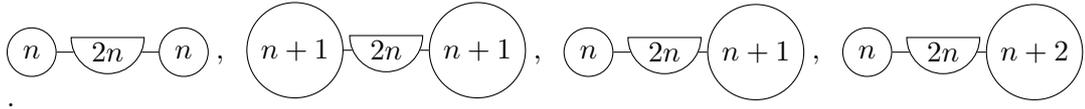
\begin{tikzpicture}
\node[SU] (A) at  (1,0)  {$n$};
\node[Sp] (B) at  (2,0)  {$2n$};
\node[SU] (C) at  (3,0)  {$n$};
\draw (A) --(B)--(C);
\end{tikzpicture}\ , \:
\begin{tikzpicture}
\node[SU] (A) at  (1,0)  {$n+1$};
\node[Sp] (B) at  (2.2,0)  {$2n$};
\node[SU] (C) at  (3.4,0)  {$n+1$};
\draw (A) --(B)--(C);
\end{tikzpicture}\ , \:
\begin{tikzpicture}
\node[SU] (A) at  (1,0)  {$n$};
\node[Sp] (B) at  (2,0)  {$2n$};
\node[SU] (C) at  (3.2,0)  {$n+1$};
\draw (A) --(B)--(C);
\end{tikzpicture}\ , \:
\begin{tikzpicture}
\node[SU] (A) at  (1,0)  {$n$};
\node[Sp] (B) at  (2,0)  {$2n$};
\node[SU] (C) at  (3.2,0)  {$n+2$};
\draw (A) --(B)--(C);
\end{tikzpicture}\ .

Some of these (those having a $\SU(n+1)$ or $\SU(n+2)$) can be extended for some values of $n$. These are all of finite A type. These can have $\SU$ branch decorations too.

\item We can take a $\USp(n)-\SU(n_1)$ branch along with a $\USp(n)-\SO(n_2)$ branch with branch currents $I_1$ and $I_2$ respectively, such that $I_1+I_2\geq0$. All such configurations give all the allowed graphs in this category. We can have further branching along the $\SO-\USp$ direction. All such graphs are of finite A or D type. There can be various decorations.

\item We can take two $\USp-\SO$ branches. Then, all such graphs can be viewed as a degenerate case of the $\USp-\SO$ trunk with more than one nodes, and can be obtained by shrinking the trunk to exactly one $\USp$ node. All these graphs are of finite A, finite D or affine D type. There can be various decorations.
\item Finally, we have one case with large length-1 branch, classified in Sec.~\ref{what?}. The only possibility here is \[
\begin{tikzpicture}
\node[1gon] (A) at  (0,0)  {$\asymT$};
\node[Sp] (B) at  (1.5,0)  {8};
\node[SO] (C) at  (3,0)  {6};
\draw (A)--(B)--(C);
\end{tikzpicture}\ .
\]
\end{itemize}

\paragraph{1-valent:} These are precisely the branches with usual current $I>0$.

\paragraph{Comment:} Notice that the only graphs in which we could take the labelling of every vertex to be arbitrarily large were found to be of finite and affine Dynkin type. The labellings on all the non-Dynkin graphs are bounded and such graphs are finite in number.

\subsubsection{$\mathsf{G}_2$ trunk}\label{g2-trunk}
\paragraph{$\mathsf{G}_2$-$\USp(2)$ branch}
There is only one: \[
 \begin{tikzpicture}[scale=.7]
\node (C) at (4,0) {$\mathsf{G}_2$};
\node[Sp] (D) at (6,0) {$2$};
\draw (C)--(D);
\end{tikzpicture}\ .
\] Note that as always in this paper, the node $\USp(2)$ is assumed to have additional one $\half\fund$ to make the node superconformal. This makes the node free of Witten's global anomaly, too.

\paragraph{$\mathsf{G}_2$-$\USp(4)$ branch}
They can all be obtained by taking $\SO(7)$-$\USp(4)$ branches and replacing $\SO(7)$ with $\mathsf{G}_2$.
Let us list some of them for illustration: 
\[
\begin{array}{l}
\begin{tikzpicture}[scale=.7]
\node (C) at (4,0) {$\mathsf{G}_2$};
\node[Sp] (D) at (6,0) {$4$};
\node[SO] (E) at (8,0) {$5$};
\node[Sp] (F) at (10,0) {$2$};
\node[SO] (G) at (12,0) {$3$};
\draw (C)--(D);
\draw[->] (D)--(E);
\draw[->] (E)--(F);
\draw[->] (F)--(G);
\end{tikzpicture}\ , \\
\begin{tikzpicture}[scale=.7]
\node (C) at (4,0) {$\mathsf{G}_2$};
\node[Sp] (D) at (6,0) {$4$};
\node[SO] (E) at (8,0) {$4$};
\draw (C)--(D);
\draw[->] (D)--(E);
\end{tikzpicture}\ ,\\
\begin{tikzpicture}[scale=.7]
\node (C) at (4,0) {$\mathsf{G}_2$};
\node[Sp] (D) at (6,0) {$4$};
\node[Sp] (E) at (8,0) {$2$};
\draw (C)--(D);
\draw[->] (D)--(E);
\end{tikzpicture}\ .
\end{array}
\]

Any $\mathsf{G}_2$-$\USp(4)$ branch can be combined with one other $\mathsf{G}_2$-$\USp(4)$ branch or two $\mathsf{G}_2$-$\USp(2)$ branches. Any $\mathsf{G}_2$-$\USp(2)$ branch can be combined with one or two or three other $\mathsf{G}_2$-$\USp(4)$ branches. A $\mathsf{G}_2$-$\USp(2)$ branch can also be combined with one other $\mathsf{G}_2$-$\USp(2)$ branch and one $\mathsf{G}_2$-$\USp(4)$ branch.

\subsection{Status of Seiberg-Witten solutions}\label{statusX}
The Seiberg-Witten solutions are known for various subcases:
\begin{itemize}
\item A well-studied class of theories  are the ones with only $\SU$ gauge nodes, $\SU$-$\SU$ bifundamentals, and $\SU$ fundamentals.
A theory in this class is often called an $\cN{=}2$ quiver gauge theory. 
As was recalled above, the associated graph is necessarily one of finite or affine Dynkin diagrams.
A universal method to obtain the Seiberg-Witten solutions of any theory of this class was obtained in \cite{Nekrasov:2012xe}; also see \cite{Katz:1997eq}.
\item A-type $\SU$ quivers were first solved in \cite{Witten:1997sc}, and D-type $\SU$ quivers were solved in \cite{Kapustin:1998xn}, both using branes.
\item A linear $\SU$ quiver theory can be    decorated by  adding symmetric or anti-symmetric two-index tensors, or an $\SO$ or $\USp$ groups at the end.  The Seiberg-Witten curves for these were found in \cite{Landsteiner:1997ei,Landsteiner:1998pb,Argyres:2002xc}. 
\item Linear $\SU$ quivers, with or without decorations by anti-symmetric two-index tensors or by $\USp$ groups at the ends, can be solved using the  construction of \cite{Gaiotto:2009we,Nanopoulos:2009xe}, using the properties of 6d $\cN{=}(2,0)$ theory of type A.   Our decomposition of the associated graph of general $\cN{=}2$ theories into a trunk and a number of branches is partially inspired by their construction. This construction was further extended in \cite{Chacaltana:2010ks,Tachikawa:2011yr,Chacaltana:2012ch} to include decorations by three-index anti-symmetric tensors of $\SU(6)$, or by branching  into D-type Dynkin diagrams, or with a twist in the loop of $\SU$ groups. 
\item Linear $\SO$-$\USp$ quivers without any decorations were first solved in \cite{Evans:1997hk,Landsteiner:1997vd,Brandhuber:1997cc}, using branes and  orientifolds. 
\item Their construction was revisited in \cite{Tachikawa:2009rb,Tachikawa:2010vg,Chacaltana:2011ze,Chacaltana:2013oka} from the point of view  of 6d $\cN{=}(2,0)$ theory of type D. This allows us to solve theories with spinors of $\SO(8)$ and $\SO(7)$, or with identifications $\SO(6)\simeq \SU(4)$, or with $\mathsf{G}_2$ gauge groups. It is hard to say at this stage exactly which subclass of the $\cN{=}2$ gauge theories classified in this paper can be solved with the methods in those papers. One natural guess is that any theory with a trunk longer than just one node, or a degenerate verion of such theory with a trunk with one node, is solvable with these methods.  Whether this is really the case remains to be seen. 
\item The Seiberg-Witten solutions to  the theories whose associated graph is of non-Dynkin type,
or those to the $\SO$-$\USp$ theories whose associated graph is of type E, are unknown so far. 
\end{itemize}

\section{Summary}\label{summary}
Let us summarize the result of the classification of $\cN{=}2$ gauge theories in this paper.
Any asymptotically free gauge theory can be obtained by removing a number of fundamental hypermultiplets from a superconformal gauge theory, see Sec.~\ref{reduction}.
Therefore, we only have to classify superconformal gauge theories. 

First, they are classified  into one of the following three possibilities:
 \begin{itemize}
\item a theory whose  gauge group is simple, discussed in Sec.~\ref{single}, or
\item a theory whose gauge group is $\SU(2)^n$ and whose hypermultiplets are given by trifundamentals, first introduced by Gaiotto \cite{Gaiotto:2009we}. This was discussed in Sec.~\ref{gaiotto}, or
\item everything else, discussed in Sec.~\ref{hardpart}.
\end{itemize}

Second, the theories in the last category have the gauge group of the form $\prod_i G_i$,
where $G_i$ is either a simple gauge group or $\SO(4)$, 
and any irreducible component of the hypermultiplets is charged under at most two of them.
Its associated graph is formed by nodes representing $G_i$,  the edges connecting two nodes, and additional `1-gons' representing non-fundamental hypermultiplets. 
Any such theory is composed of a trunk and branches, as shown in Sec.~\ref{overall} and enumerated in Sec.~\ref{branches}.

Given a theory, the graph thus formed consists of either 
\begin{itemize}
\item a single loop, treated in Sec.~\ref{loop},
\item a trunk with more than one node, plus small branches attached to the ends, discussed in Sec.~\ref{tree-with-a-long-trunk},
\item a trunk with a single node, analyzed in detail in Sec.~\ref{su-trunk}, Sec.~\ref{so-trunk}, Sec.~\ref{usp-trunk}, Sec.~\ref{g2-trunk} for $\SU$, $\SO$, $\USp$ and $\mathsf{G}_2$ respectively, or
\item three very rare cases listed in Sec.~\ref{veryrare}.
\end{itemize}

We also indicated, along the way, which of the theories have been solved in the sense of Seiberg and Witten. Interested readers are urged to solve the remaining ones.

\section*{Acknowledgements}
 L.B. would like to thank Alok Laddha, Anirban Basu and Anudhyan Boral for useful discussions.  L.B. is also extremely grateful to  Sujay Ashok for making this project possible. 
 Y.T. would like to thank Abhijit Gadde and Wenbin Yan for informing the authors of a series of previous papers \cite{Koh:1983ir,Dong:1984vt,Derendinger:1984bu,Jiang:1984wr,Jiang:1984nb,Jiang:1984wn,Jiang:1984aa} concerning a partial classification.
 The authors would like to thank Jacques Distler for carefully reading the manuscript and making helpful suggestions concerning the known Seiberg-Witten solutions. 
 They would also like to thank Pierre Fayet, Hirosi Ooguri, and Edward Witten for informing them a few points to be corrected concerning the historical developments in an earlier version of the manuscript. 
 Finally, the authors thank Antonio Sciarappa and Gabi Zafrir for pointing out that the last two entries of Sec. \ref{Antonio} were missing from previous versions of the manuscript. The last entry there was also missing from Table \ref{1gon-isolated} which lists all the allowed representations.

 The work of L.B. is supported in part by the University of Tokyo Research Internship Program (UTRIP) for undergraduates. The work of Y.T. is supported  in part by JSPS Grant-in-Aid for Scientific Research No. 25870159 and in part by WPI Initiative, MEXT, Japan at IPMU, the University of Tokyo.

\bibliographystyle{ytphys}
\let\bbb\bibitem\def\bibitem{\itemsep4pt\bbb}
\bibliography{ref}

\end{document}